\documentclass[12pt]{article}
\usepackage{amsmath}
\usepackage{graphicx}
\usepackage{enumerate}
\usepackage{natbib}
\usepackage{url} 

\usepackage{physics}
\usepackage{xcolor}
\usepackage{amssymb}
\usepackage{amsthm}
\usepackage{multirow}
\usepackage{float}

\newtheorem{lemma}{Lemma}

\newcommand{\argmin}[1]{ \underset{#1}{\textup{argmin}} }

\newcommand{\blind}{0}

\begin{document}

\def\spacingset#1{\renewcommand{\baselinestretch}%
{#1}\small\normalsize} \spacingset{1}


\begin{center}
{\LARGE\bf Beyond time-homogeneity for continuous-time multistate Markov models}
\end{center}

\if0\blind
{
\begin{center}
Emmett B. Kendall$^1$, Jonathan P. Williams$^{1,2}$, Gudmund H. Hermansen$^{2,3,4}$, \\
Frederic Bois$^5$, and Vo Hong Thanh$^5$ 
\end{center}

\begin{center}
\small
$^1$Department of Statistics, North Carolina State University\\
$^2$Centre for Advanced Study, Norwegian Academy of Science and Letters\\
$^3$University of Oslo\\
$^4$Peace Research Institute Oslo (PRIO)\\
$^5$CERTARA UK Limited, Simcyp Division, Level 2-Acero
\end{center}
} \fi

\bigskip
\begin{abstract}
Multistate Markov models are a canonical parametric approach for data modeling of observed or latent stochastic processes supported on a finite state space.  Continuous-time Markov processes describe data that are observed irregularly over time, as is often the case in longitudinal medical data, for example.  Assuming that a continuous-time Markov process is time-homogeneous, a closed-form likelihood function can be derived from the Kolmogorov forward equations -- a system of differential equations with a well-known matrix-exponential solution.  Unfortunately, however, the forward equations do not admit an analytical solution for continuous-time, time-{\em inhomogeneous} Markov processes, and so researchers and practitioners often make the simplifying assumption that the process is piecewise time-homogeneous.  In this paper, we provide intuitions and illustrations of the potential biases for parameter estimation that may ensue in the more realistic scenario that the piecewise-homogeneous assumption is violated, and we advocate for a solution for likelihood computation in a truly time-inhomogeneous fashion.  Particular focus is afforded to the context of multistate Markov models that allow for state label misclassifications, which applies more broadly to hidden Markov models (HMMs), and Bayesian computations bypass the necessity for computationally demanding numerical gradient approximations for obtaining maximum likelihood estimates (MLEs). Supplemental materials are available online.
\end{abstract}

\noindent%
\vfill

\newpage
\spacingset{1.5} 

\section{Introduction}\label{sec:intro}

Given a physical process that evolves over time, it is not always reasonable to expect that the {\em rate of evolution} remains constant over time.  An obvious example of such a process is the progression of dementia as it develops with age.  The rate at which young individuals (e.g., less than 50 years of age) develop dementia is virtually null compared to the rate observed for individuals beyond 70 years of age \citep[i.e., all else constant, the probability of developing dementia within one year for an individual of age 40 is most certainly different than that probability for an individual of age 80;][]{williams2020bayesian}.  Fitting a continuous-time multistate Markov model or HMM in a time-inhomogeneous fashion, however, (i.e., where the transition rate matrix is a continuous function of time) has been a crux of numerous methodological approaches in the statistics community for many decades \citep[e.g.,][]{faddy1976note, kay1986markov, Jackson2003, titman2011flexible, Jack2016, williams2020bayesian, simeon2021multistate}.  

It is well-established that a continuous-time Markov process is characterized as the solution to the system of differential equations commonly referred to as the Kolmogorov forward equations \citep[see, e.g.,][]{karlin1981second}.  While in the time-homogeneous case a closed-form matrix-exponential solution to the Kolmogorov forward equations exists, no known analytical solution is available in the time-inhomogeneous case.  Accordingly, an analytical expression for the likelihood function of a time-inhomogeneous multistate Markov model is unavailable, and so a variety of approximations have been proposed.  For example, as in \cite{williams2020bayesian}, if it is assumed that the continuous-time Markov process transition rates are constant between integer ages, then the closed-form matrix-exponential solution to the forward equations exists for every integer age.  Thus, in the piecewise time-homogeneous case, the full likelihood can be expressed with the transition probability matrices computed as products of multiple matrix-exponential solutions indexed by integer age.  This solution facilitates the identifiability of time-inhomogeneous transition rates insofar as they are strictly piecewise-constant functions of time, which is surely an unrealistic, albeit pragmatic assumption about the biological process.  Although pragmatic, approximating transition rates that are continuous functions of time, by rates that are piecewise-constant functions of time, of finer and finer resolution (e.g., yearly, monthly, weekly, etc.), quickly becomes computationally infeasible due to the rapidly growing number of matrix-exponential calculations required at finer resolutions.

Our contributions are: first, to provide intuitions and illustrations of the potential estimation biases that might result from modeling transition rates as piecewise-constant functions of time, when they are truly continuous functions of time; second, to advocate for numerical ordinary differential equation (ODE) solutions to the time-inhomogeneous Kolmogorov forward equations that correctly specifies the transition rates as continuous functions of time; and third, to demonstrate, with both real and synthetic data sets, instances of when differences may and may not arise in estimated parameter values between the continuous versus piecewise-constant rates approaches -- indicating the practical significance of numerical ODE solutions for fitting continuous-time, time-inhomogeneous Markov processes (HMMs or otherwise).  Moreover, due to the computational complexity of matrix-exponential calculations, we find that a numerical ODE solution -- which we implement via the \texttt{R} package \texttt{deSolve} \citep[][]{soetaert2010package} -- leads to reduced computational time, even for time-homogeneous processes.  The numerical ODE solver approach allows for computing the likelihood, but it does not yield a closed-form expression of the likelihood function.  As such, gradient-based algorithms for finding MLEs of model parameters may be computationally prohibitive because the gradient might need to be approximated numerically.  In any case, we propose Markov chain Monte Carlo (MCMC) computations within a fully Bayesian paradigm, avoiding the need to compute first derivatives of the likelihood function.

Markov models and HMMs play a fundamental role in many recent applications from modeling disease progression \citep{williams2020bayesian} and seizure detection \citep{furui2021timeseries}, to assisting in volcanic eruption risk evaluation \citep{aspinall2006using}, as well as studying the efficacy of ceasefires on conflict dynamics \citep{williams2021ceasefires}.  Applying numerical ODE solvers to fit continuous-time Markov models (HMMs or otherwise) in a truly time-inhomogeneous fashion has broad implications, particularly for studying longitudinal medical and biological data sets where the underlying phenomena exhibit continuously time-evolving rates of progression.

An earlier approach contributed by \cite{titman2011flexible} proposed numerical methods for solving the Kolmogorov forward equations in the context of continuous-time, time-inhomogeneous multistate Markov processes, but does not consider HMMs.  In \cite{titman2011flexible}, the transition rates are assumed to be smooth functions of time, and with the canonical strategy of using B-splines to model the transition rates as functions of time, Fisher scoring is used to compute MLEs of the model parameters.  Derivatives with respect to the transition rate parameters are calculated numerically by extending the system of forward equations.  While there exists a thorough investigation into the computational efficiency of applying numerical approaches, little is said in \cite{titman2011flexible} about the inferential consequences of using these numerical methods as opposed to more common, approximation-based estimation procedures in the multistate Markov modeling literature. Other ideas to circumvent the piecewise-homogeneous approximation include time transformation techniques applied to the transition rates based on a common, smooth function of time that makes calculating MLEs more feasible \citep{hubbard2008modeling}, and penalized splines of individual rates in a forward-only model \citep{joly2002penalized}.  Further, there exist alternatives to the HMM framework for specification and estimation of time-dependent rates of progression, such as stochastic simulation algorithms \citep{thanh2015simulation, marchetti2017simulation}.

More recently, the methods of \cite{titman2011flexible} have been extended to accommodate the HMM framework in the \texttt{R} package \texttt{nhm} \citep{titmanPackage}, and for some model specifications first derivatives of the likelihood function can be computed analytically.  Bayesian implementations are not yet available in the \texttt{nhm} package.  In addition to investigating a Bayesian approach, our paper provides further clarity, intuition, and empirical studies to illustrate the consequences of a piecewise-homogeneous approximation when transition rates may be continuous functions of time.  Hence, our paper goes beyond the scope of what is presented in \cite{titman2011flexible} by facilitating broader awareness for implementing these numerical ODE solvers within the context of continuous-time, time-inhomogeneous Markov models, as well as applying them to more complex data applications that go beyond the purview of what is capable of \texttt{nhm}. Additionally, \cite{simeon2021multistate} points to the lack of these numerical-integration techniques being applied in the statistics literature for correctly fitting multistate Markov models (HMMs or otherwise).  Furthermore, \cite{simeon2021multistate} argues that piecewise-homogeneous approximations are slow and inaccurate relative to numerical methods.

Another alternative to solving the time dependent Kolmogorov forward equations is to empirically estimate the transition probability matrix via the product-integral solution to the Kolmogorov forward equations (see \cite{borgan1997three}, \cite{aalen2008survival}, and \cite{andersen2012statistical} for more detail). An extension of the Kaplan-Meier estimator \citep{kaplan1958nonparametric}, the Aalen-Johansen estimator \citep{aalen1978empirical} is an empirical product limit estimator for the transition probability matrix of a finite state, inhomogeneous Markov process. Although the Aalen-Johansen estimator offers a nonparametric approach to estimating the transition probability matrix, the empirical calculation of the estimator assumes that all state transitions are observed and observation times correspond to state transition times \citep{borgan1997three, aalen2008survival}. Hence, similar to the piecewise-homogeneous approach, the Aalen-Johansen estimator relies on making assumptions about the evolution of the state process which, in most applications, are likely not met and can thus result in estimation bias (as shown in Sections \ref{sec:simulation} and \ref{subsec:cavReal}). That being stated, an appeal of this estimator is that it can serve as a nonparametric MLE for the true transition probability matrix, and under certain conditions, is a uniformly consistent estimator  \citep{andersen2012statistical}. The Aalen-Johansen estimator is a function of the Nelson-Aalen estimator \citep{nelson1969hazard, nelson1972theory}, a nonparametric estimator of the cumulative transition intensity of the time-inhomogeneous Markov process. The Aalen-Johansen and Nelson-Aalen estimators are frequently used for assessing model goodness-of-fit when implementing other approaches to fitting multistate Markov models \citep{bureau2003applications, titman2008general, titman2010model, soper2020hidden}. Extending beyond Markov processes in particular, much work has been done to explore asymptotic properties of the Aalen-Johansen estimator in non-Markov settings \citep[e.g.,][]{datta2001validity, datta2002estimation, glidden2002robust, gunnes2007estimating}. 


The last alternative to solving the time dependent Kolmogorov forward equations we consider is a discretization of time, altogether.  A detailed comparison of discrete-time versus continuous-time, time-homogeneous models can be found in \cite{wan2016comparison}.  They find that if inter-observation times are similar, a discrete-time model is as effective as a continuous-time model, as well as computationally more efficient. See also, \cite{aalen1997markov} and \cite{liu2015efficient} for discretized approaches for fitting continuous-time Markov models.  We present evidence, in Section \ref{mice_data}, suggesting the reasonableness of discrete-time approximations for time-inhomogeneous HMMs -- when regularly-spaced observation times are available at a high resolution -- based on our analysis of real electrocorticography (ECoG) data recorded from a mice sleep study \cite[the data are from][]{bojarskaite2020}.  Nonetheless, the importance of our investigation of continuous-time processes is built on the fact that many data sets do {\em not} exhibit observation times near enough a regular spacing that a discrete-time approximation would be effective.

The remainder of the paper is structured as follows. In Section \ref{sec:methods}, the multistate Markov model and HMM frameworks are introduced, and mathematical intuition is provided for the potential biases of a piecewise-homogeneous approximation for continuously evolving time-inhomogeneous transition rates in the context of a simple linear regression illustration. Section \ref{sec:emp_results} covers all details for the data analyses beginning with data descriptions (Section \ref{subsec:dat_descr}) and modeling approaches (Section \ref{subsec:compete}). Then, Section \ref{sec:simulation} presents a simulation study of synthetic data generated from a continuous-time, time-inhomogeneous Markov process, to compare fitting an HMM with piecewise-homogeneous approximations, empirical approximations via the Aalen-Johansen and Nelson-Aalen estimators, as well as the time-inhomogeneous numerical ODE solution.  We consider comparisons of these approaches on real data sets in Section \ref{sec:real_data}.  The paper ends with concluding remarks in Section \ref{sec:conclusion}.  The \texttt{R} code along with a workflow script file for reproducing all numerical results presented in this paper is available at \if0\blind{\url{https://ebkendall.github.io/research.html}}\fi \if1\blind{\texttt{blinded}}\fi.

\section{Multistate Markov model methods and inference}\label{sec:methods}

We first formalize the common notations and specifications for multistate models. Let $\vb*{s_i} = \{s_{i,1}, \dots, s_{i,n_{i}}\}$, for $i \in \{1,\dots,n\}$ independent, represent an \textit{observed} state sequence from a discrete, multistate process, $X(t)$, where for any time $t\geq 0$,  $X(t) \in \{1,\dots, m\}$, for some fixed, positive integer $m$. The inferential goal with regard to these $n$ independent, multistate processes is to learn parameters characterizing the transitions between the $m$ states over time. If $t_{i,1} < t_{i,2} < \dots < t_{i,n_{i}}$, for $i \in \{1,\dots,n\}$, represents the observed time points for each $\vb*{s_i}$, then the joint probability mass function is given by
\begin{align*}
f_{s_{1},\dots,s_{n}}(\vb*{s_1}, \dots, \vb*{s_n}) & = \prod_{i=1}^{n} P(X(t_{i,1}) = s_{i,1}) \cdot \prod_{k=2}^{n_i}P\qty(X(t_{i,k}) = s_{i,k} \mid \{X(t_{i,j}) = s_{i,j}\}_{j=1}^{k-1}).
\end{align*}
For the remainder of this paper, however, we make the stochastic assumption that the evolution of $X(t)$ is defined by a Markov process. In particular, the Markov assumption simplifies the expression for $f_{s_{1},\dots,s_{n}}(\vb*{s_1}, \dots, \vb*{s_n})$ as follows.  For any $t,h \geq 0$, let $\vb*{P}(t, t+h)$ be the $m\times m$ transition probability matrix over the time period from $t$ to $t+h$, for $X(t)$, where the $r^{\text{th}}$ row and $s^{\text{th}}$ column of $\vb*{P}(t, t+h)$, denoted $P_{r,s}(t, t+h)$, represents the probability of transitioning from state $r$ to state $s$ over the time period $t$ to $t+h$. The Markov assumption thus allows the joint probability mass function to be expressed as
\begin{equation} \label{eq:like1}
f_{s_{1},\dots,s_{n}}(\vb*{s_1}, \dots, \vb*{s_n}) = \prod_{i=1}^{n} P(X(t_{i,1}) = s_{i,1}) \cdot P_{s_{i,1},s_{i,2}}(t_{i,1},t_{i,2}) \cdots P_{s_{i,n_{i}-1},s_{i,n_{i}}}(t_{i,n_{i}-1},t_{i,n_{i}}),
\end{equation}
where
$$P_{s_{i,k-1},s_{i,k}}(t_{i,k-1},t_{i,k}) = P\{X(t_{i,k}) = s_{i,k} \mid X(t_{i,k-1}) = s_{i,k-1} \},$$
for $k = 2,3,\dots,n_i$. In what follows, we detail the various approaches for defining $\vb*{P}(t, t+h)$ depending on further stochastic assumptions made about the evolution of $X(t)$.


The mathematical expression for $\vb*{P}(t, t+h)$ is determined by whether time is measured as discrete or continuous.  While the discrete-time approach leads to a simple expression (i.e., multiplying the one-time-point-ahead transition probability matrix $h$ times, where $t$ and $h$ take values from a discrete grid of points such as the nonnegative integers), it is not adequate for modeling data that are observed irregularly in time.  Thus, it is the necessary complexities of the continuous-time approach that we focus on in this paper.

In the continuous-time, time-homogeneous setting, $t$ and $h$ take nonnegative, real values, and $\vb*{P}(t,t+h)$ is equivalent to $\vb*{P}(0,h)$.  Further, so long as the transition probabilities $P_{r,s}(0,h)$, for $r,s \in \{1,\dots,m\}$, are differentiable functions of $h$, $\vb*{P}(0,h)$ is defined by an $m\times m$ {\em infinitesimal generator} or {\em transition rate} matrix, labeled $\vb*{Q}$ (i.e., the infinitesimal analogue of the one-time-point-ahead transition probability matrix arising in the discrete-time setting), via the system of differential equations
\begin{equation}\label{eq:kolm}
\vb*{P}'(0,h) = \vb*{P}(0,h) \cdot \vb*{Q},
\end{equation}
commonly referred to as the {\em Kolmogorov forward equations}.  The $(r,s)^{\text{th}}$ component of $\vb*{Q}$, denoted $q_{r,s}$, can be interpreted as the instantaneous rate of transition from state $r$ to state $s \neq r$:
\begin{equation}\label{eq:qComp}
q_{r,s} := \lim_{\delta \downarrow 0} \frac{P\{X(\delta) = s \mid X(0) = r)\}}{\delta},
\end{equation}
and $q_{r,r} := -\sum_{s : s\neq r} q_{r,s}$ for $r,s \in \{1,\dots,m\}$ \citep[see, e.g.,][for comprehensive mathematical details]{karlin1981second}.  By standard matrix calculus arguments, it is readily verifiable that equation (\ref{eq:kolm}) has the analytical matrix-exponential solution
\[
\vb*{P}(0,h) = e^{h \cdot \vb*{Q}}.
\]
This closed-form solution makes it uncomplicated to express (\ref{eq:like1}) as
\begin{align*}
f_{s_{1},\dots,s_{n}}(\vb*{s_1}, \dots, \vb*{s_n} \mid \vb*{Q}) & = \prod_{i=1}^{n} P(s_{i,1}) \cdot P_{s_{i,1},s_{i,2}}(0,t_{i,2}-t_{i,1}) \cdots P_{s_{i,n_{i}-1},s_{i,n_{i}}}(0,t_{i,n_{i}}-t_{i,n_{i}-1}) \\
& = \prod_{i=1}^{n} P(s_{i,1}) \cdot \{e^{(t_{i,2}-t_{i,1}) \cdot \vb*{Q}}\}_{s_{i,1},s_{i,2}} \cdots \{e^{(t_{i,n_{i}}-t_{i,n_{i}-1}) \cdot \vb*{Q}}\}_{s_{i,n_{i}-1},s_{i,n_{i}}},
\end{align*}
where $P(s_{i,1})$ is the initial state probability, and $\{\cdot\}_{r,s}$ denotes the $(r,s)^{\text{th}}$ component of the matrix argument.  Moreover, this likelihood expression admits a closed-form gradient expression with respect to the components of $\vb*{Q}$, leading to efficient optimization algorithms for computing the MLEs, and the tractability of both the likelihood and its gradient are easily extended to the HMM framework for dealing with latent state sequences.  

Statistical inferences are typically motivated by characterizing the components of $\vb*{Q}$ as a function of covariates; for example, it may be of interest to determine if transition rates over a discretized progression of a disease (i.e., over $m$ states of severity) are affected by baseline covariates such as sex, race, biomarkers, socioeconomic status, etc.  It is not possible, however, to consider non-baseline covariates -- such as age -- that may be important for describing key features of the data (e.g., when studying dementia progression or other aging-related processes).  This is due to the fact that the matrix-exponential solution to the Kolmogorov forward equations (\ref{eq:kolm}) relies on the assumption that $\vb*{Q}$ is constant over time; this assumption is also why it suffices to take $t = 0$ in equation (\ref{eq:kolm}).  Nonetheless, it remains paramount in many scientific and machine learning applications to develop extensions for estimating the effects of time-varying covariates on the transition rates in a continuous-time, time-inhomogeneous multistate Markov process defined by
\begin{equation}\label{eq:inhomo_kolm}
\vb*{P}'(t, t+h) = \vb*{P}(t, t+h) \cdot \vb*{Q}(t+h),
\end{equation}
where $\vb*{Q}(t+h)$ is the transition rate matrix defined as a function of time $t > 0$. The components of $\vb*{Q}(t+h)$ are reminiscent of (\ref{eq:qComp}), except are now functions of time, given by
\[
q_{r,s}(t+h) := \lim_{\delta \downarrow 0} \frac{P\{X(t+h+\delta) = s \mid X(t+h) = r)\}}{\delta},
\]
and $q_{r,r}(t+h) := -\sum_{s : s\neq r} q_{r,s}(t+h)$ for $r,s \in \{1,\dots,m\}$; see also \cite{karlin1981second, titman2010model} and \cite{titman2011flexible}.

Unfortunately, there exists no known analytical solution to the time-inhomogeneous forward equations (\ref{eq:inhomo_kolm}) -- at least to the best of our knowledge -- and rather than relying on numerical ODE solutions for likelihood and gradient computations, a common strategy observed throughout the statistical literature is to approximate a time-inhomogeneous Markov process by one that is piecewise time-homogeneous.  In that case, the closed-form matrix-exponential solution applies in a piecewise-constant fashion, so that the Kolmogorov forward equations can be expressed as
\begin{equation}\label{eq:inhomo_kolm_piecewise}
\vb*{P}'(t, t+h) = \vb*{P}(t, t+h) \cdot \vb*{Q}(g_{d}(t)),
\end{equation}
where $g_{d}(t) := t - (t \mod d)$ for some fixed $d > 0$.  In the case $d = 1$ it follows that $g_{1}(t) = \lfloor t \rfloor$, i.e., the {\em floor} function taking the greatest integer less than or equal to $t$ as its value.  So, for example, if $d = 1$ and $t$ is time as measured by the age of an individual, then $g_{1}(t) = \lfloor t \rfloor$ is the integer-valued age, and so defining $\vb*{Q}$ as a function of integer age means that the transition rates only update with every birthday of the individual.  Accordingly, in this example, for any $t \ge 0$ and $h \ge 0$ equation (\ref{eq:inhomo_kolm_piecewise}) has the closed-form solution 
\[
\vb*{P}(t,t+h) =
\begin{cases}
e^{(\lfloor t + 1\rfloor - t) \cdot \vb*{Q}(\lfloor t \rfloor)} \cdot e^{1 \cdot \vb*{Q}(\lfloor t +1 \rfloor)} \cdot e^{1 \cdot \vb*{Q}(\lfloor t +2 \rfloor)}  \cdots e^{(t+h - \lfloor t+h\rfloor) \cdot \vb*{Q}(\lfloor t+h \rfloor)} & \text{if } \lfloor t \rfloor \ne \lfloor t+h \rfloor \\
e^{h \cdot \vb*{Q}(\lfloor t+h \rfloor)} & \text{else.} \\
\end{cases}
\]
In general, for arbitrary $d > 0$, the solution to equation (\ref{eq:inhomo_kolm_piecewise}) is
\[
\vb*{P}(t,t+h) =
\begin{cases} 
e^{( g_{d}(t + d) - t) \cdot \vb*{Q}(g_{d}(t))} \cdot e^{d \cdot \vb*{Q}(g_{d}(t+d))} \cdots  e^{(t+h - g_{d}(t+h)) \cdot \vb*{Q}(g_{d}(t+h))} & \text{if } g_{d}(t) \ne g_{d}(t+h) \\
e^{h \cdot \vb*{Q}(g_{d}(t+h))} & \text{else.} \\
\end{cases}
\]
It is seemingly a reasonable and pragmatic strategy to approximate the solution to the time-inhomogeneous forward equations (\ref{eq:inhomo_kolm}) by choosing sufficiently small $d > 0$, but bias on the transition rate estimates may persist, nonetheless -- if the transition rates of the underlying process are truly evolving in continuous-time -- and the matrix-exponential computations involved in the piecewise solution can have a tremendous impact on computation time for small $d > 0$.  Moreover, numerical ODE solutions exist for directly solving the system (\ref{eq:inhomo_kolm}) and are readily available in standard software such as the \texttt{R} packages \texttt{deSolve} \citep[][]{soetaert2010package} and \texttt{nhm} \citep[][]{titmanPackage}.   

To provide insight and intuition on the biases on the transition rate parameters that may ensue from approximating a continuous-time process by a piecewise-constant function, we provide a simple illustration for how such biases arise in a linear model context next, in Section \ref{sec:simple_illustration}.

\subsection{Simple illustration of piecewise-constant approximation bias}\label{sec:simple_illustration}

To understand how estimation biases on the transition rate parameters $\vb*{Q} = \vb*{Q}(t)$ may result from solving the system (\ref{eq:inhomo_kolm_piecewise}) with $\vb*{Q} = \vb*{Q}(g_{d}(t))$, it is insightful to understand how biases will result from fitting the parameters of a simple linear model depending on continuous-time $t$ by a linear model depending on piecewise-constant-time $g_{d}(t) = t - (t \mod d)$ for some fixed $d > 0$.  Accordingly, suppose that $y_{1}, \dots, y_{n}$ are real-valued observations on the line
\[
y_{i} = \beta_{0} + \beta_{1} t_{i},
\]
where $t_{i}$ denotes an instance of a value in continuous-time, for $i \in \{1,\dots,n\}$, and $\beta_{0}$ and $\beta_{1}$ are fixed scalars.  Then regressing $y_{1}, \dots, y_{n}$ on the piecewise-constant-time instances $g_{d}(t_{1}), \dots g_{d}(t_{n})$ yields the least squares projection $b_{0,n} + b_{1,n} g_{d}(t_{i})$, for $i \in \{1,\dots,n\}$, where
\begin{equation}\label{eq:projection_coefs}
\begin{split}
b_{0,n} & = \beta_{0} + \frac{\overline{t}_{n}\{\sum_{i=1}^{n}g^{2}_d(t_{i})\} - \frac{1}{n}\sum_{i,j}t_{i}\cdot g_{d}(t_{i})g_{d}(t_{j}) }{ A_{n} } \beta_{1} \\
b_{1,n} & = \beta_{1} + \bigg(\frac{ \sum_{i=1}^{n}(t_{i} - \overline{t}_{n})\{g_{d}(t_{i}) - \frac{1}{n}\sum_{k=1}^{n}g_{d}(t_{k})\} }{ A_{n} } - 1\bigg) \beta_{1},
\end{split}
\end{equation}
$\overline{t}_{n} := \frac{1}{n}\sum_{i=1}^{n} t_{i}$, and assuming $A_{n} := \sum_{i=1}^{n}\{g_{d}(t_{i}) - \frac{1}{n}\sum_{k=1}^{n}g_{d}(t_{k})\}^{2} > 0$.  

\begin{figure}[H]
\spacingset{1}
\centering
\includegraphics[scale=.34, trim={0mm 6mm 9mm 0mm}, clip]{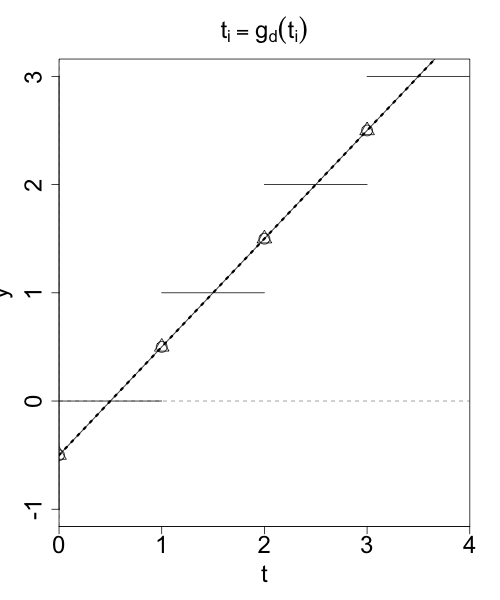}
\includegraphics[scale=.34, trim={18mm 6mm 9mm 0mm}, clip]{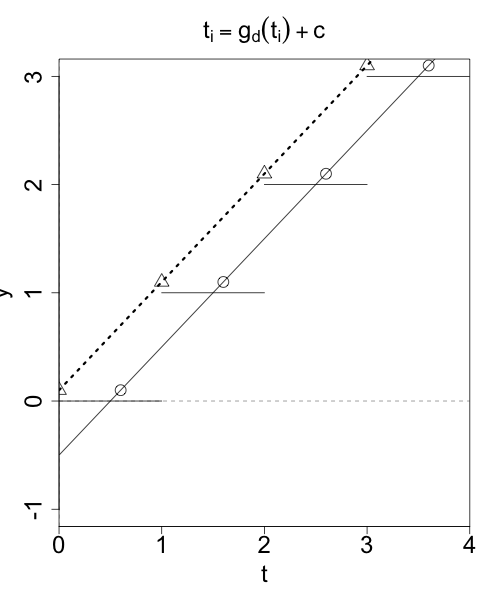}
\includegraphics[scale=.34, trim={18mm 6mm 9mm 0mm}, clip]{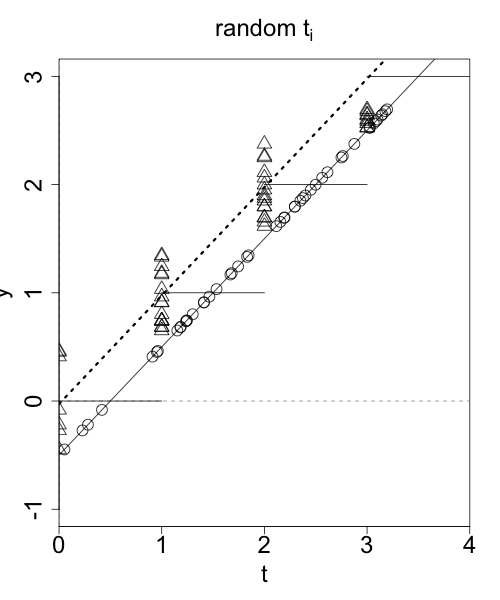}
\vspace{-.1in}
\caption{\footnotesize Illustration of Lemmas \ref{lemma:slope} and \ref{lemma:baseline} with $d = 1$, i.e., $g_{d}(t) = \lfloor t \rfloor$, for synthetic values $y_{i} = -\frac{1}{2} + t_{i}$, for $i \in \{1,\dots,n\}$, for time instances $t_{1},\dots,t_{n}$.  The left panel corresponds to integer-valued time instances $t_{i} = \lfloor t_{i} \rfloor$ (i.e., $c = 0$, as in Lemma \ref{lemma:baseline}), the middle panel corresponds to shifted time instances $t_{i} = \lfloor t_{i} \rfloor + c$ (with $c = .6$, as in Lemma \ref{lemma:baseline}), and the right panel corresponds to randomly generated time instances $t_{i}$.  The solid black lines are plots of the line $-\frac{1}{2} + t$, and the dashed black lines are plots of the fitted least squares projection lines $b_{0,n} + b_{1,n}t$.  The circles denote scatter plots of the observed points $\{( t_{i}, y_{i})\}_{i=1}^{n}$, whereas the triangles denote scatter plots of the points $\{( g_{d}(t_{i}), y_{i})\}_{i=1}^{n} = \{( \lfloor t_{i} \rfloor, y_{i})\}_{i=1}^{n}$ -- used to fit the dashed projection lines.}\label{fig:simple_illustration}
\end{figure}

For fixed $d > 0$, so long as sufficient variation is observed in the $g_{d}(t_{1}), \dots g_{d}(t_{n})$ (i.e., there is an unbounded and heavy-tailed distribution of available time instances $t_{1}, \dots, t_{n}$, for large $n$), the bias $\beta_{1} - b_{1,n}$ for the slope parameter will vanish for large $n$, as stated in Lemma \ref{lemma:slope}.  The bias $\beta_{0} - b_{0,n}$ for the baseline parameter, however, need not vanish for large $n$, as illustrated by Lemma \ref{lemma:baseline}.  Figure \ref{fig:simple_illustration} presents a graphical representation of these two lemmas.

\begin{lemma}\label{lemma:slope}
For fixed $d > 0$ and for every positive integer $n$, the least squares projection coefficients $b_{0,n}$ and $b_{1,n}$ can be expressed as in (\ref{eq:projection_coefs}), and if $A_{n} > 0$, then
\[
|\beta_{1} - b_{1,n}| \le \frac{d \cdot |\beta_{1}|}{(\frac{1}{n}A_{n})^{\frac{1}{2}}}.
\]
In particular, if $\lim_{n\to\infty}\frac{1}{n}A_{n} = \infty$, then $\lim_{n\to\infty} b_{1,n} = \beta_{1}$.
\end{lemma}
\begin{proof}
See Section \ref{sec:proofs}.
\end{proof}

\begin{lemma}\label{lemma:baseline}
For fixed $d > 0$, suppose that $t_{i} = g_{d}(t_{i}) + c$, for some $c \ge 0$, for all $i \in \{1,\dots,n\}$. Then, for every positive integer $n$ with $A_{n} > 0$,
\[
|\beta_{0} - b_{0,n}| = c \cdot |\beta_{1}|.
\]
In particular, in the case $\beta_{1} \ne 0$, it follows that $b_{0,n} = \beta_{0}$ if and only if the observed time instances $t_{1}, \dots, t_{n}$ happen to coincide precisely at discontinuity points of the function $g_{d}(\cdot)$.
\end{lemma}
\begin{proof}
See Section \ref{sec:proofs}.
\end{proof}

As illustrated by the panels in Figure \ref{fig:simple_illustration}, projecting points from a line that is continuous in time onto a piecewise-constant function of time can result in substantial bias in fitting the baseline parameter.  Although the transition probability matrix for a continuous-time, time-inhomogeneous Markov process is more complicated than a simple linear function of time, the simple illustrations presented in this section provide the intuition for how fitting the rate parameters based on a piecewise-homogeneous rate matrix approximation $\vb*{Q}(g_{d}(t))$ will likely result in biased estimation of the baseline rates of $\vb*{Q}(t)$.  

In Section \ref{sec:simulation}, we provide empirical evidence to demonstrate such baseline rate bias using synthetic data generated from continuous-time, time-inhomogeneous Markov processes. Furthermore, we demonstrate, in Section \ref{subsec:cavReal}, indications that the baseline rate bias can manifest from real data by comparing rate parameter estimates from fitting a continuously evolving time-inhomogeneous Markov model -- based on a numerical ODE solution to (\ref{eq:inhomo_kolm}) -- versus a piecewise-homogeneous Markov model -- based on the matrix-exponential-based solution to (\ref{eq:inhomo_kolm_piecewise}).  We also present a second real data analysis in Section \ref{mice_data} suggesting that the piecewise-homogeneous approximation may be reasonable {\em if} the observed time instances fall on a grid, as described by $c = 0$ in Lemma \ref{lemma:baseline}.

\subsection{Hidden Markov models}\label{sec:hmm}
In order to evince the broadness of scope in which our analyses and conclusions apply, we further extend the continuous-time Markov model framework with the alternative assumption that the state sequences $\vb*{s_i} = \{s_{i,1}, \dots, s_{i,n_{i}}\}$, for $i \in \{1,\dots,n\}$, are \textit{unobserved}. The state sequences $\vb*{s_i}$ are still assumed to be defined by a continuous-time, time-inhomogeneous Markov process; however, now let $\vb*{y_{i}} = \{y_{i,1}, \dots, y_{i,n_{i}}\}$ represent \textit{observed} response data, where the standard HMM formulation assumes that their random variable analogues $Y_{i,1}, \dots, Y_{i,n_{i}}$ are conditionally independent given the unobserved $s_{i,1}, \dots, s_{i,n_{i}}$, and that $Y_{i,k}$ only depends on $s_{i,k}$, for every $k \in \{1,\dots,n_{i}\}$, for every $i \in \{1,\dots,n\}$. 
The same issues and potential rate biases -- as discussed for (non-latent) Markov models -- apply for approximating a time-inhomogeneous HMM by a piecewise-homogeneous HMM, but the likelihood function must be modified to account for uncertainty in the latent state sequences and the observed response data.

Given the standard HMM formulation, the density for $\vb*{y_{i}}$ can be derived by integrating over all possible unobserved state sequences: 
\begin{align*}
f_{y_{i}}(\vb*{y_i}) & =  \sum_{s_{i,1}=1}^{m} \cdots \sum_{s_{i, n_{i}}=1}^{m} f_{y_{i}\mid s_{i}}(\vb*{y_i} \mid \vb*{s_i}) \cdot f_{s_{i}}(\vb*{s_i}) = \sum_{s_{i,1}=1}^{m} \cdots \sum_{s_{i, n_{i}}=1}^{m} f_{s_{i}}(\vb*{s_i}) \cdot \prod_{k=1}^{n_{i}} f_{y_{i,k}\mid s_{i,k}}(y_{i,k} \mid s_{i,k}),
\end{align*}
where, as defined by the continuous-time Markov process that solves the time-inhomogeneous forward equations (\ref{eq:inhomo_kolm}),
\[
f_{s_{i}}(\vb*{s_i}) = P(s_{i,1}) \cdot \prod_{k=2}^{n_{i}} P_{s_{i,k-1},s_{i,k}}(t_{i,k-1},t_{i,k}).
\]
Accordingly,
\begin{align*}
f_{y_{i}}(\vb*{y_i}) & = \sum_{s_{i,1}=1}^{m} P(s_{i,1}) \cdot f_{y_{i,1}\mid s_{i,1}}(y_{i,1} \mid s_{i,1}) \cdot \sum_{s_{i,2}=1}^{m}P_{s_{i,1},s_{i,2}}(t_{i,1},t_{i,2}) \cdot f_{y_{i,2}\mid s_{i,2}}(y_{i,2}\mid s_{i,2}) \cdots \\
& \hspace{1in} \times \sum_{s_{i,n_{i}}=1}^{m} P_{s_{i,n_{i}-1},s_{i,n_{i}}}(t_{i,n_{i}-1},t_{i,n_{i}}) \cdot f_{y_{i,n_{i}}\mid s_{i,n_{i}}}(y_{i,n_{i}}\mid s_{i,n_{i}}),
\end{align*}
which yields a joint density function over independent sequences $\vb*{y_{1}},\dots,\vb*{y_{n}}$:
\begin{equation}\label{eq:likelihood}
f_{y_{1},\dots,y_{n}}(\vb*{y_{1}},\dots,\vb*{y_{n}}) = \prod_{i=1}^{n}\vb*{\pi}^{T} \vb*{D}_{(i,1)} \cdot \vb*{P}(t_{i,1}, t_{i,2}) \vb*{D}_{(i,2)} \cdots \vb*{P}(t_{i,n_{i}-1}, t_{i,n_{i}}) \vb*{D}_{(i,n_{i})} \cdot \vb*{1}_{m},
\end{equation}
where $\vb*{\pi} := \{P(s_{i,1} = 1),\dots,P(s_{i,1} = m)\}^{T}$, 
\[
\spacingset{1}
\vb*{D}_{(i,k)} :=
\begin{pmatrix}
f_{y_{i,k}\mid s_{i,k}}(y_{i,k}\mid s_{i,k} = 1) & & \\
& \ddots & \\
& & f_{y_{i,k}\mid s_{i,k}}(y_{i,k}\mid s_{i,k} = m)
\end{pmatrix},
\]
and $\vb*{1}_{m}$ is an $m\times1$ vector with each component taking the value 1.  As per the discussions in Section \ref{sec:methods}, this likelihood function can only be computed numerically if $\vb*{P}(t,t+h)$ is defined by the time-inhomogeneous forward equations (\ref{eq:inhomo_kolm}), but it admits a closed-form approximation (for some fixed $d > 0$) based on the piecewise-homogeneous forward equations (\ref{eq:inhomo_kolm_piecewise}).  These are the likelihood expressions that we compare using synthetic and real data in the sections that follow.

\section{Empirical results}\label{sec:emp_results}
\subsection{Data descriptions}\label{subsec:dat_descr}
 
Three data sets are studied in Sections \ref{sec:simulation} and \ref{sec:real_data}: real cardiac allograft vasculopathy (CAV) data (Section \ref{subsubsec:cavDesc}), synthetic CAV data (Section \ref{subsec:simDataDesc}), and real ECoG mice sleep study data (Section \ref{subsec:eeg_mice}). Each data application serves a unique purpose in illustrating the practicality of using numerical ODE solvers for evaluating the transition probability matrix versus relying upon the piecewise time-homogeneous approximation or the Aalen-Johansen estimator, in the context of continuous-time, time-inhomogeneous Markov models. The focus is on comparing the resulting inferences from using the aforementioned modeling approaches, and specifically how the estimation of the transition rate parameters differ.

Although common in the statistical literature to use the piecewise time-homogeneous assumption, if, for the real data analyses, the resulting parameter estimates are different from that of the numerical ODE approach, then the decision to use one method over the other is not to be overlooked. This is the motivation for the real data analyses because it is unknown which approach ``truly'' describes the process being modeled; therefore, if the inferential conclusions are different, practitioners must consciously decide if the piecewise-homogeneity assumption accurately represents the process, otherwise, the numerical ODE solver eliminates the need for that assumption. The same applies for implementing the Aalen-Johansen estimator and the assumption of observing all state transition times.  More broadly, the choice to use the CAV data is both because it is well-studied in the multistate Markov model literature, as well as it offers data with unequal inter-observation times within subjects; alternatively, the ECoG mice sleep study provides an example where the inter-observation times within and between subjects is the same.  We generate synthetic CAV data in the context of the simulation study to illustrate how fitting the rate parameters of a continuous-time, time-inhomogeneous Markov process, based on a piecewise-homogeneous approximation -- with rate matrix $\vb*{Q}(g_{d}(t))$ -- can lead to biased estimation of the baseline rates of $\vb*{Q}(t)$. 

\subsubsection{Cardiac allograft vasculopathy study data} \label{subsubsec:cavDesc}
The first of the real data considered is the CAV data, available in the \texttt{msm} package for \texttt{R} \citep{jackson2011multi}, which comprises information and clinical assessments collected from 622 subjects who underwent heart transplant surgery.  Data for each subject are available on roughly an annual basis for the years following surgery, and classification of the severity of deterioration of the subjects' arterial walls are recorded based on the clinical assessments \citep[i.e., severity of CAV][]{jackson2011multi}.  Associated with each clinical visit, for each subject, is the severity of CAV, classified as either 1: ``no CAV''; 2: ``mild/moderate CAV''; 3: ``severe CAV''; or 4: ``death.''  In the context of a multistate Markov process, these classifications define four states.  Time must be treated as continuous -- due to the irregular observation times of the clinical visits -- and the observed state labels for CAV severity, associated with each clinical visit for each subject, must be assumed observed-with-error to account for diagnostic/clerical uncertainty.

\subsubsection{Simulated data}\label{subsec:simDataDesc}
The synthetic data is generated from a time-inhomogeneous Markov process based on HMM parameters fit to the well-studied CAV data set described in Section \ref{subsubsec:cavDesc}. In particular, the evolution of the Markov process rate matrix for this simulation is constructed to be inhomogeneous in continuous-time. Additionally, observations are made at irregular time instances, similar to the real CAV data, thus not guaranteeing all state transitions are observed, nor do the observation times necessarily correspond to transition times. Therefore, when the continuous-time, time-inhomogeneous HMM is fit to the synthetic data using either the piecewise-homogeneous approximation or the Aalen-Johansen estimator, compared to the numerical ODE solution approach, we are studying the model when it is both misspecified and correctly specified, respectively.

\subsubsection{Mice sleep study data}\label{subsec:eeg_mice}
In the context of data collected from controlled experiments, especially when processing of the data is required, it is common that observations are recorded on a regularly-spaced grid of time instances.  Thus, for the second real data application, ECoG data from a mice sleep study are used in which inter-observation times are equal. This study pertains to biological processes for which it is believed unrealistic to assume that transition rates are constant over time \citep{bojarskaite2020}.  ECoG and electromyography (EMG) measurements are recorded for each mouse at a frequency of 5,000 observations per second, over hours.  We construct an HMM to infer state sequences over the sleep-states 1: ``intermediate state (IS)''; 2: ``non-rapid eye movement (NREM)''; or 3: ``rapid eye movement (REM).''  Sleep researchers have manually annotated state labels that are provided in the data set, but not at all observed time instances and we assume that misclassifications are possible.  Hence, the theorized true state sequences are regarded as latent, and the observed state labels are understood as a response variable.  We also use the observed ECoG time series as an additional response variable.  In other sleep study applications of HMMs, researchers have considered the sleep-state labels as observed-without-error \citep[e.g., ][]{doroshenkov2007classification,pan2012transition,ghimatgar2019automatic}, or have attempted to label sleep-states in an unsupervised fashion \citep[e.g., ][]{flexerand2002automatic}.

The full details for the preprocessing of the raw data are provided in Section 2 of \cite{crainiceanu2009nonparametric}; however, instead of splitting the data into 30-second epochs and applying the fast Fourier transform (FFT), we use 5-second epochs.  Additionally, we thinned the data to  a frequency of approximately 556 observations per second instead of 5,000.  Once the FFT is applied on all epochs of the ECoG time series, what remains is a sequence of complex numbers corresponding to distinct frequencies.  The modulus of FFT-mapped values determines the frequencies of waves that comprise the ECoG signal.  In the context of studying sleep, the focus is on four separate frequency bands of brain waves and how the patterns they exhibit are signatures of the sleep-states.  These wave bands and their respective frequency ranges (in Hz) are $\delta: [.8, 4.0], \theta: [4.2, 8.0], \alpha: [8.2, 13.0],$ and $\beta: [13.2, 20.0]$ \citep{crainiceanu2009nonparametric}.  For every 5-second epoch, each of the four wave bands has an associated power, and we record their relative powers.  Accordingly, the data set we analyze is discretized in 5-second time increments, and to each time instance is an associated relative power of the $\delta$, $\theta$, $\alpha$, and $\beta$ bands, along with an observed sleep-state label, if available.  If more than one observed sleep-state label resides in a single 5-second epoch, for a given mouse, then the state label is considered missing (i.e., the likelihood function integrates over all states for that epoch, for that mouse).

\subsection{Competing methods}\label{subsec:compete}
The five modeling approaches that will be compared are:
\begin{description}
\item[(A) :] The numerical ODE solution (computed using the \texttt{deSolve} package \citep[][]{soetaert2010package} for \texttt{R}) to the correctly specified time-inhomogeneous forward equations (\ref{eq:inhomo_kolm}), via a custom Metropolis-Hastings random walk MCMC algorithm.
\item[(B) :] The numerical ODE solution (computed using the \texttt{nhm} package \citep[][]{titmanPackage} for \texttt{R}) to the correctly specified time-inhomogeneous forward equations (\ref{eq:inhomo_kolm}), via MLE.
\item[(C) :] The closed-form matrix-exponential-based solution to the time-homogeneous approximation (\ref{eq:inhomo_kolm_piecewise}), via a custom  Metropolis-Hastings random walk MCMC algorithm.  
\item[(D) :] The closed-form matrix-exponential-based solution to the time-homogeneous approximation (\ref{eq:inhomo_kolm_piecewise}), via MLE (computed using the \texttt{msm} package \citep[][]{jackson2011multi} for \texttt{R}).
\item[(E) :] The empirical approximation to the transition rate and transition probability matrices using the Nelson-Aalen and Aalen-Johansen estimators, respectively (computed using the \texttt{AalenJohansen} package \citep{bladt2023conditional} for \texttt{R}).
\end{description}
Again, the primary interest for comparison is in how the inference differs depending on the approach used to evaluate the transition probability matrix for continuous-time, time-inhomogeneous Markov processes.  Details of the properties of maximum likelihood estimation for continuous-time, multistate Markov processes are provided in \cite{kalbfleisch1985analysis, kay1986markov}, and others. A similar discussion but for Bayesian MCMC approaches is found in \cite{zhao2016bayesian}. See  \cite{moler2003nineteen} for methods to evaluate the matrix exponential in (C) and (D). Lastly, for more theory related specifically to HMMs, \cite{leroux1992maximum} details consistency results for maximum likelihood estimation and \cite{vernet2015posterior} comments on posterior consistency.

\subsubsection{Remarks on computation}
For the Bayesian approaches (A) and (C), the MCMC sampling is defined by a Metropolis-Hastings random walk. An adaptive proposal strategy is used during the burnin period to learn the proposal covariance for better mixing and quicker convergence. For approaches (B) and (D), maximum likelihood estimation is done via the BHHH \citep{titmanPackage} and BFGS \citep{jackson2011multi} algorithms, respectively. Both our custom numerical approach, (A), and the numerical approach in \texttt{nhm}, (B), implement the differential equation solvers from \texttt{deSolve} \citep{soetaert2010package} which utilize the efficient and adaptive integration scheme developed in \cite{petzold1983automatic}. Note that when applying the \texttt{nhm} package, upper bounds on the parameter space are needed and successive observations with a time gap less than .001 are manually increased by .001 for numerical stability purposes. Lastly, in all data applications in Sections \ref{sec:simulation} and \ref{sec:real_data}, we assume a parametric form for the transition rate matrix. Note that the Aalen-Johansen and Nelson-Aalen estimators are not designed to be applied in the modeling setups of Sections \ref{sec:simulation} and \ref{sec:real_data} because the inferential focus for those analyses is to fit a given parametric form of the transition rates. Nonetheless, although the estimators for approach (E) are nonparametric, Section \ref{sec:ajEst} provides the details of a method we constructed for associating the transition rate parameters with the Nelson-Aalen estimator over all time points (a verification of the efficacy of this approach is provided in the supplemental materials). 
It is important to highlight that the \texttt{AalenJohansen} package offers the \texttt{prodint} function which numerically computes the product-integral representation of the transition probability matrix, analogous to \texttt{deSolve} with Riemann integration. Thus, a possible alternative to approach (A) would be to use \texttt{prodint} instead of \texttt{deSolve}; however, a more thorough application of approach (A) with \texttt{prodint} is not conducted because it would have been computationally infeasible given the MCMC structure and the size of the data used (for more information, see the supplemental materials). That being said, based on preliminary investigations, approach (A) with either \texttt{prodint} or \texttt{deSolve} should lead to similar results.

\subsection{Simulation study}\label{sec:simulation}
Recall that the synthetic data is based on the CAV data from the \texttt{msm} package. Let the observed state label, $Y_{i,k}$, for the $k^{\text{th}}$ observation of the $i^{\text{th}}$ subject define an HMM response variable.  Assuming that only state-adjacent misclassifications are possible, the probability mass function for $Y_{i,k}$ conditional on the true state $s_{i,k}$ is given by the rows of the table:
\begin{center}
\spacingset{1}
\begin{tabular}{c c|c c c c}
& & & $y_{i,k}$ : observed state & &\\
&        & 1 & 2 & 3 & 4 \\
\hline
& 1 & $1-p_1$ & $p_1$ & 0 & 0 \\
$s_{i,k}$ : true state & 2 & $p_2$ & $1-p_2-p_3$ & $p_3$ & 0\\
& 3 & 0 & $p_4$ & $1-p_4$ & 0\\
& 4 & 0 & 0 & 0 & 1\\
\end{tabular}
\end{center}
where $p_1, p_2, p_3, p_4 \in [0,1]$ are unknown parameters \citep[see also,][]{sharples2003diagnostic}.  Next, assuming that the progression of CAV is irreversible \citep[e.g., as in][]{sharples2003diagnostic,jackson2011multi}, the transition rate matrix has the forward-model specification:
\[
\spacingset{1}
\vb*{Q}(t) = \mqty(-q_1 - q_2 & q_1 & 0 & q_2\\
                              0 & -q_3 - q_4 & q_3 & q_4\\
                              0 & 0 & -q_5 & q_5\\
                              0 & 0 & 0 & 0),
\]
and we model the rates as functions of covariates in a standard log-linear fashion \citep[][]{jackson2011multi,williams2020bayesian}:
\begin{equation}\label{eq:rates_CAV}
q_{j} = \exp(\beta_{0,j} + \beta_{1,j} \cdot t + \beta_{2,j} \cdot \text{sex}),
\end{equation}
for $j \in \{1,\dots,5\}$, where $t$ is time in years since heart transplant surgery, ``sex'' is the biological sex of the subject, and $\beta_{0,j}$, $\beta_{1,j}$, and $\beta_{2,j}$ are unknown rate-specific coefficient parameters.  This simple parametric model defines a continuous-time, time-inhomogeneous, multistate HMM that is fit to both the real and synthetic CAV data sets, and statistical inferences are made on the transition rates for the progression of CAV in the years following heart transplant surgery.  

We first fit the model to the CAV data set by implementing a Metropolis-Hastings random walk MCMC algorithm on a posterior density kernel constructed with the likelihood function (\ref{eq:likelihood}), where the transition probability matrices are computed numerically from the time-inhomogeneous forward equations (\ref{eq:inhomo_kolm}) using the ODE solver in the \texttt{deSolve} package \citep[][]{soetaert2010package} in \texttt{R}.  The prior densities for initial state probabilities $\vb*{\pi}$ and the response function parameters $p_1, p_2, p_3, p_4$ are specified as logit-transformed Gaussian densities with diffuse scale parameters, and the transition rate parameters $\beta_{0,j}$, $\beta_{1,j}$, and $\beta_{2,j}$, for $j \in \{1,\dots,5\}$, are assigned Gaussian densities with diffuse scale parameters.  The fitted posterior means are used to define the {\em true parameter values} of a continuous-time, time-inhomogeneous, multistate Markov process, from which we generate 100 synthetic data sets (though, we scale the {\em true} slope parameters $\beta_{1,j}$ by a factor of 3 for the purpose of illustration).  Each of the 100 synthetic data sets consists of 2000 sampled subjects, and inter-observation times and sex covariate values are drawn consistent with the empirical frequencies observed in the real CAV data set.

\begin{figure}[H]
\spacingset{1}
\centering
\includegraphics[scale =.237]{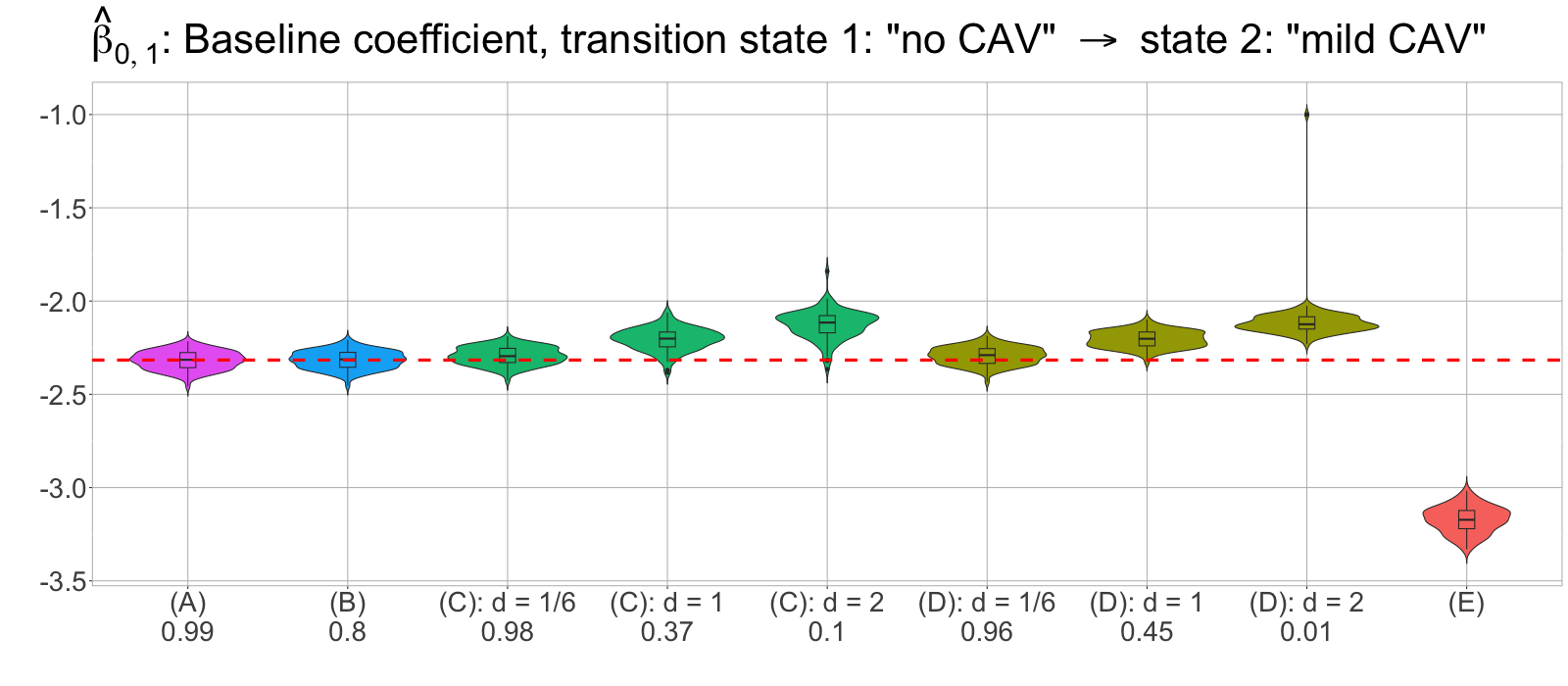}\\
\includegraphics[scale =.237]{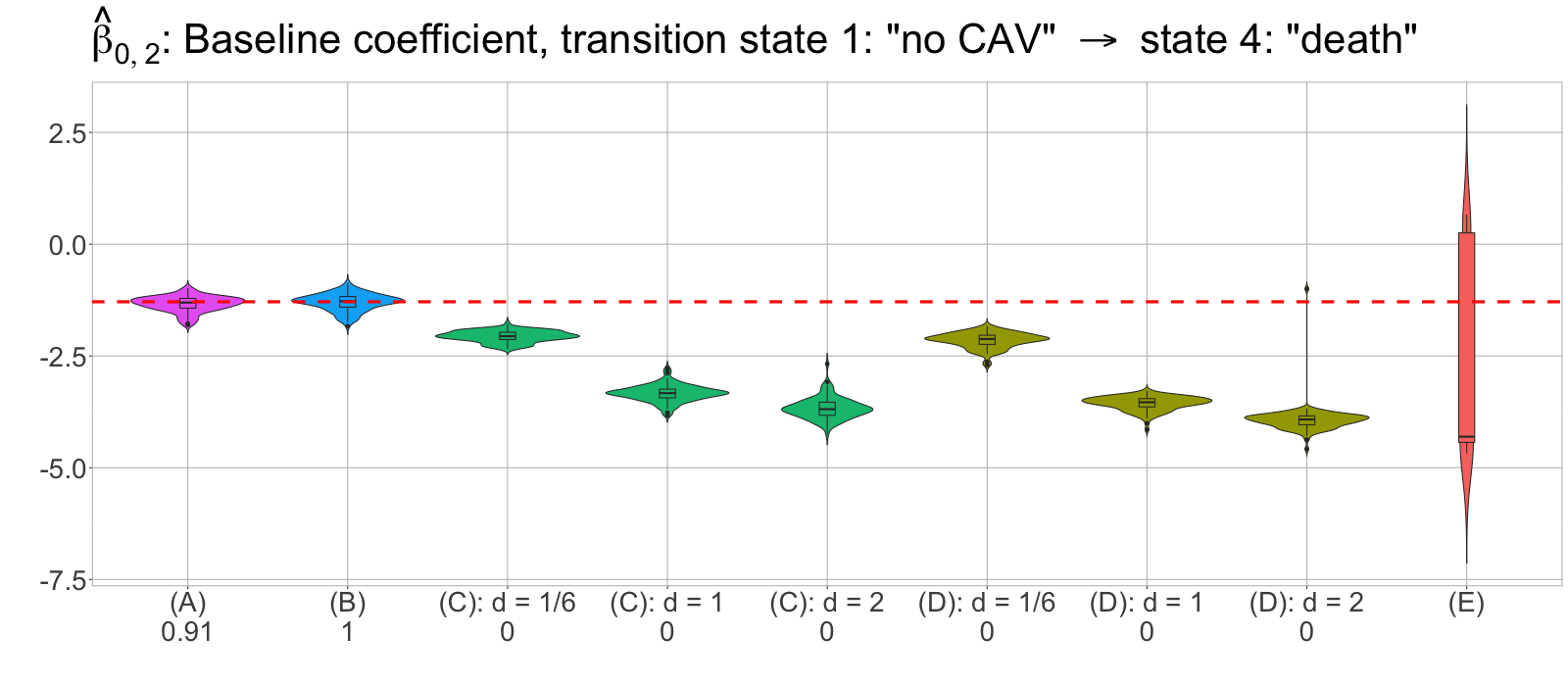}\\
\includegraphics[scale =.237]{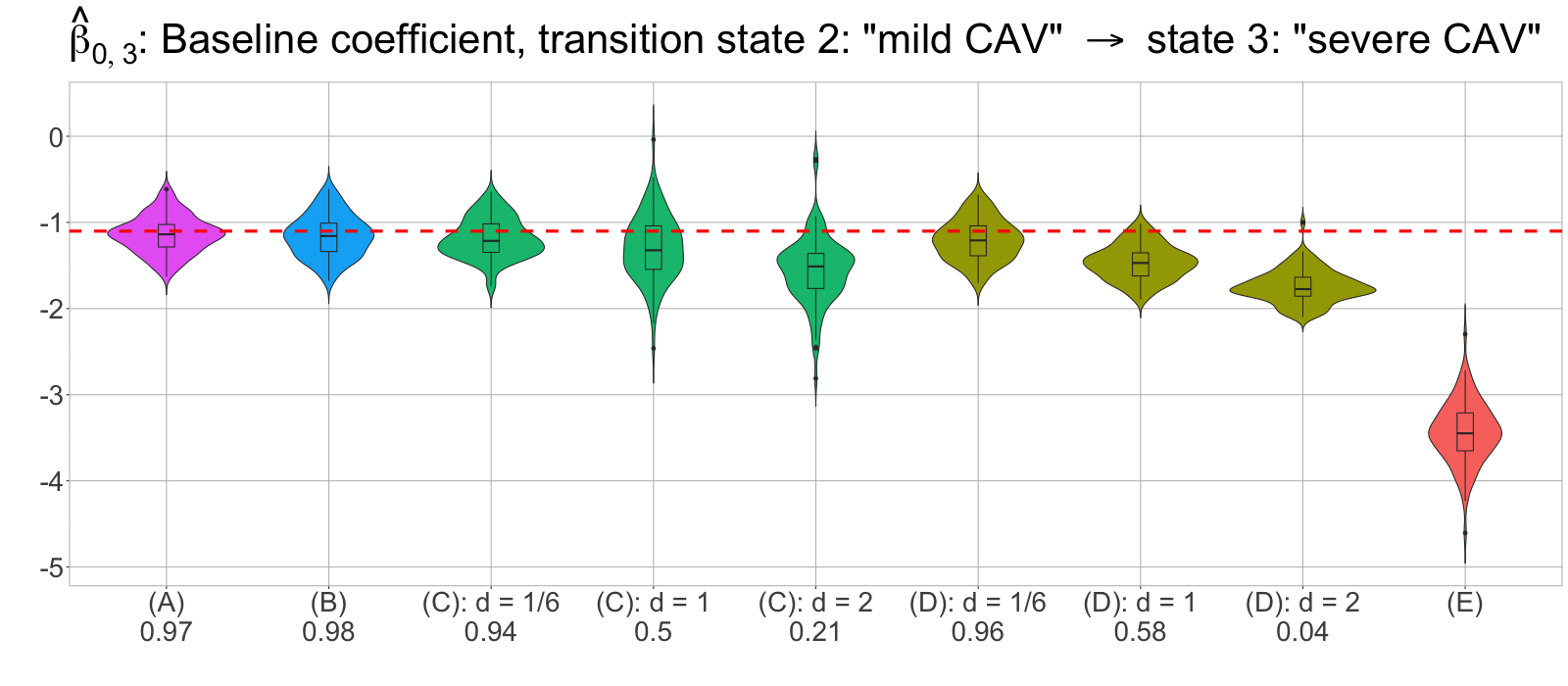}
\caption{\footnotesize For approaches (A) and (C), presented are violin plots of the posterior means corresponding to the simulation study of 100 synthetic data sets.  The violin plots associated with approaches (B) and (D) are of the MLEs.  The violin plots for approach (E) are from the empirical estimation process outlined in Section \ref{sec:ajEst}. The baseline coefficient parameters for the transition rates are as specified in equation (\ref{eq:rates_CAV}). Where available, the numbers below each label report coverages of .95 level credible or confidence intervals.}\label{fig:sim_violin1}
\end{figure}

\begin{figure}[H]
\spacingset{1}
\centering
\includegraphics[scale =.237]{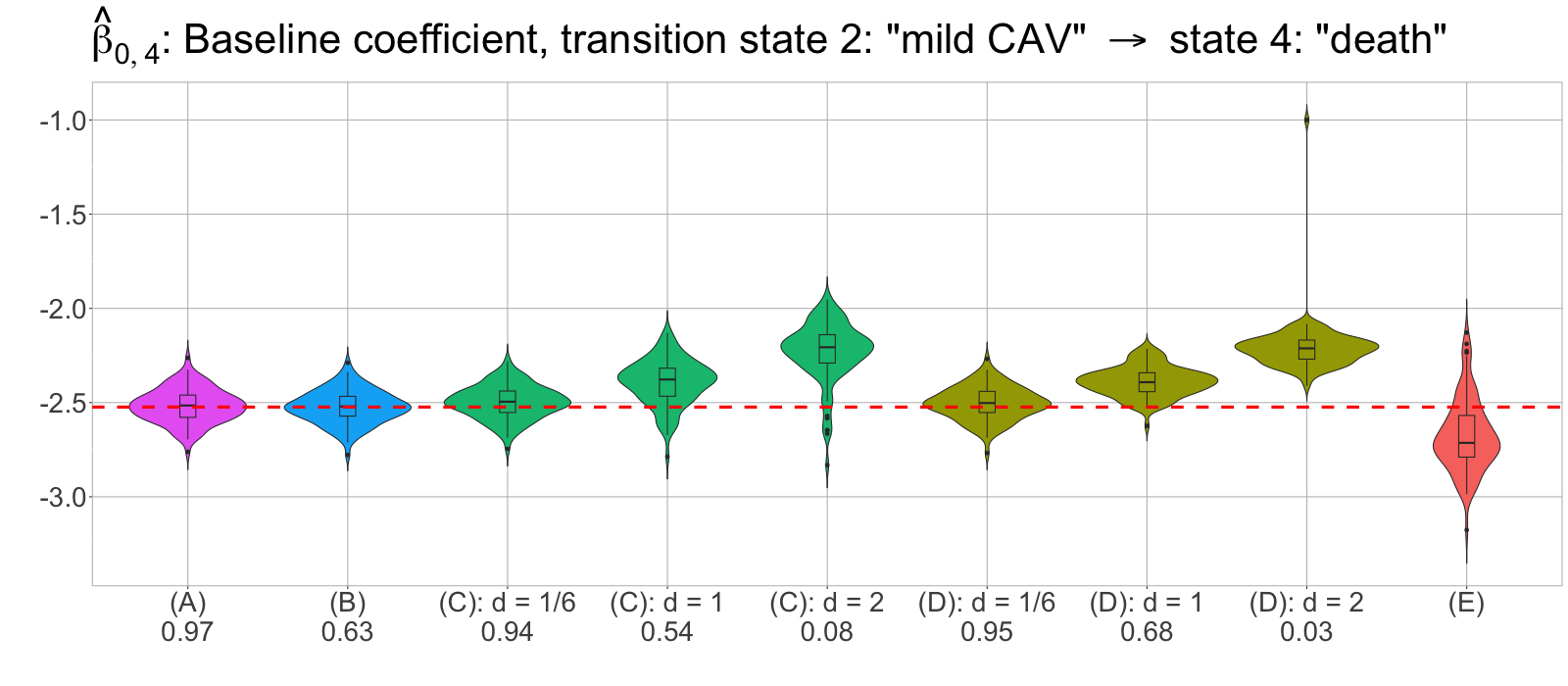}\\
\includegraphics[scale =.237]{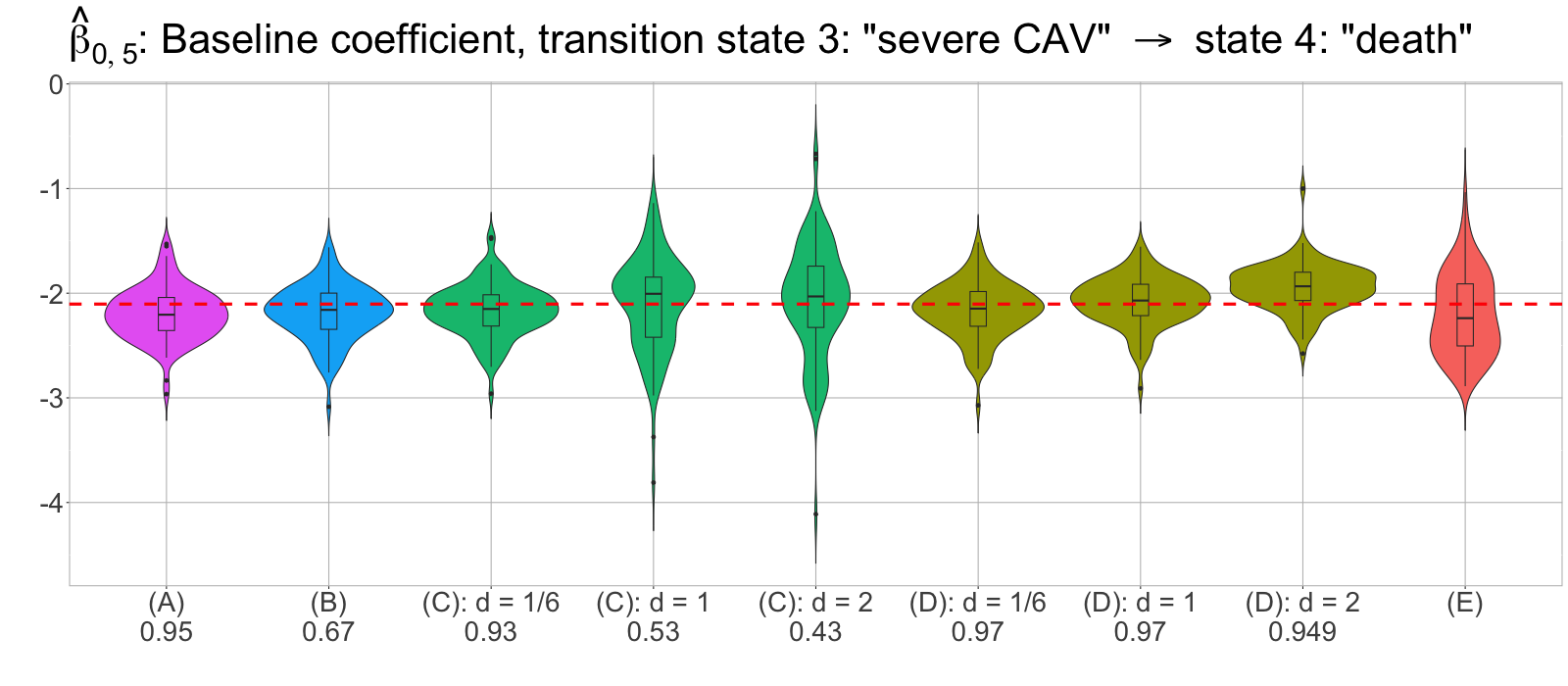}
\caption{\footnotesize  For approaches (A) and (C), presented are violin plots of the posterior means corresponding to the simulation study of 100 synthetic data sets.  The violin plots associated with approaches (B) and (D) are of the MLEs. The violin plots for approach (E) are from the empirical estimation process outlined in Section \ref{sec:ajEst}. The baseline coefficient parameters for the transition rates are as specified in equation (\ref{eq:rates_CAV}). Where available, the numbers below each label report coverages of .95 level credible or confidence intervals.}\label{fig:sim_violin2}
\end{figure}

For the simulation study, we compare the five approaches to fitting the HMM as outlined in Section \ref{subsec:compete}. Furthermore, the matrix-exponential-based solutions (C) and (D) are fitted -- for each of the 100 synthetic data sets -- for each choice of $d \in \{1/6, 1, 2\}$, as in (\ref{eq:inhomo_kolm_piecewise}), representing time-homogeneity at the resolution of bi-monthly, yearly, and bi-yearly, respectively.  The solutions (C) and (D) become computationally prohibitive at the monthly resolution, i.e., $d = 1/12$.  A summary of the results is displayed in Figures \ref{fig:sim_violin1} and \ref{fig:sim_violin2}; violin plots of the posterior means or MLEs for the baseline coefficient parameters for each transition rate are shown, along with the coverage of .95 level credible or confidence intervals.  Note that solutions (C) and (D) represent the model fit when it is misspecified because the data are simulated with the rate matrix updating continuously as a function of time, not piecewise-constant with respect to time. Next, solution (E) represents a model fit when it is misspecified for the following three reasons: states are observed with error, not all transitions between states are necessarily observed, and observation times do not necessarily correspond to transition times. Only solutions (A) and (B) are fit using a correctly specified model. Violin plots for all HMM parameters are in the supplemental materials.

Consistent with the intuition provided in Section \ref{sec:simple_illustration}, it is evident from Figures \ref{fig:sim_violin1} and \ref{fig:sim_violin2} that substantial estimation bias for the baseline rate parameters (i.e., $\beta_{0,j}$ for $j \in \{1,\dots,5\}$) is exhibited by the HMMs fitted with the time-homogeneous approximation (\ref{eq:inhomo_kolm_piecewise}), for each choice of $d \in \{1/6, 1, 2\}$, whereas the coefficient estimates based on the numerical ODE solver approaches to fitting the time-inhomogeneous HMM (\ref{eq:inhomo_kolm}) are well-concentrated on the true coefficient values. Additionally, a similar bias is observed with regards to approach (E). Note that despite approach (B) leading to unbiased estimates, the coverage probabilities for $\beta_{0,1}$, $\beta_{0,4}$, and $\beta_{0,5}$ are markedly less than the nominal level of .95, in contrast to those of (A). Moreover, less time is required for computing the likelihood function based on the numerical ODE solution than is required for the matrix-exponential-based computations for $d =1/6$, as shown in Table \ref{tab:sim_time}.  There are observed, however, reduced computation times for the matrix-exponential-based solutions for $d\in\{1,2\}$, but these come at the cost of more substantial estimation bias.

\begin{table}[H]
\centering\spacingset{1}\footnotesize
\begin{tabular}{l c c c c }
\hline
approach & (A) & (C): $d=1/6$ & (C): $d=1$  & (C): $d=2$  \\
likelihood evaluation time (seconds) & 2.53 & 2.73 & .93 & .75\\
\hline
\end{tabular}
\caption{\footnotesize Displayed are the times to compute the likelihood function evaluated at the true parameter values, averaged over the 100 synthetic data sets.}
\label{tab:sim_time}
\end{table}

\subsection{Real data results}\label{sec:real_data}

\subsubsection{Cardiac allograft vasculopathy study data}\label{subsec:cavReal}

As described in the simulation study in Section \ref{sec:simulation}, the continuously time-evolving transition rates were constructed as a feature of the synthetic or {\em artificial} CAV data (in order to assess the estimation performance observed when fitting the HMMs to data that are actually generated from a Markov process with this feature).  Instead, in this section, we model the real CAV data set in the same manner as Section \ref{sec:simulation} and compare the five approaches (A), (B), (C), (D), and (E) except now, time-evolving transition rates {\em may not} be a feature exhibited by these data, but in that case the slope coefficients for $t$ (i.e., $\beta_{1,j}$ for $j \in \{1,\dots,5\}$) would be estimated as zero and the five approaches should largely yield the same inference on the baseline rates.  Thus, any observed difference in the estimated transition rates is likely attributable to time-evolving rates being exhibited by the data, and it would be rather contrived to assume that the underlying rates of CAV progression actually evolve in a piecewise-constant fashion -- let-alone to identify the correct resolution of the discontinuity points. Similarly, there is no reason to believe that the progression of CAV occurs precisely at the observed time points. Hence, any discrepancies in the resulting inference from the five approaches is indicative that future practitioners applying continuous-time, time-inhomogeneous Markov models should make a conscious and justified decision for which method of evaluating the transition probability matrix is used.

\begin{figure}[H]
\spacingset{1}
\centering
\includegraphics[scale =.29]{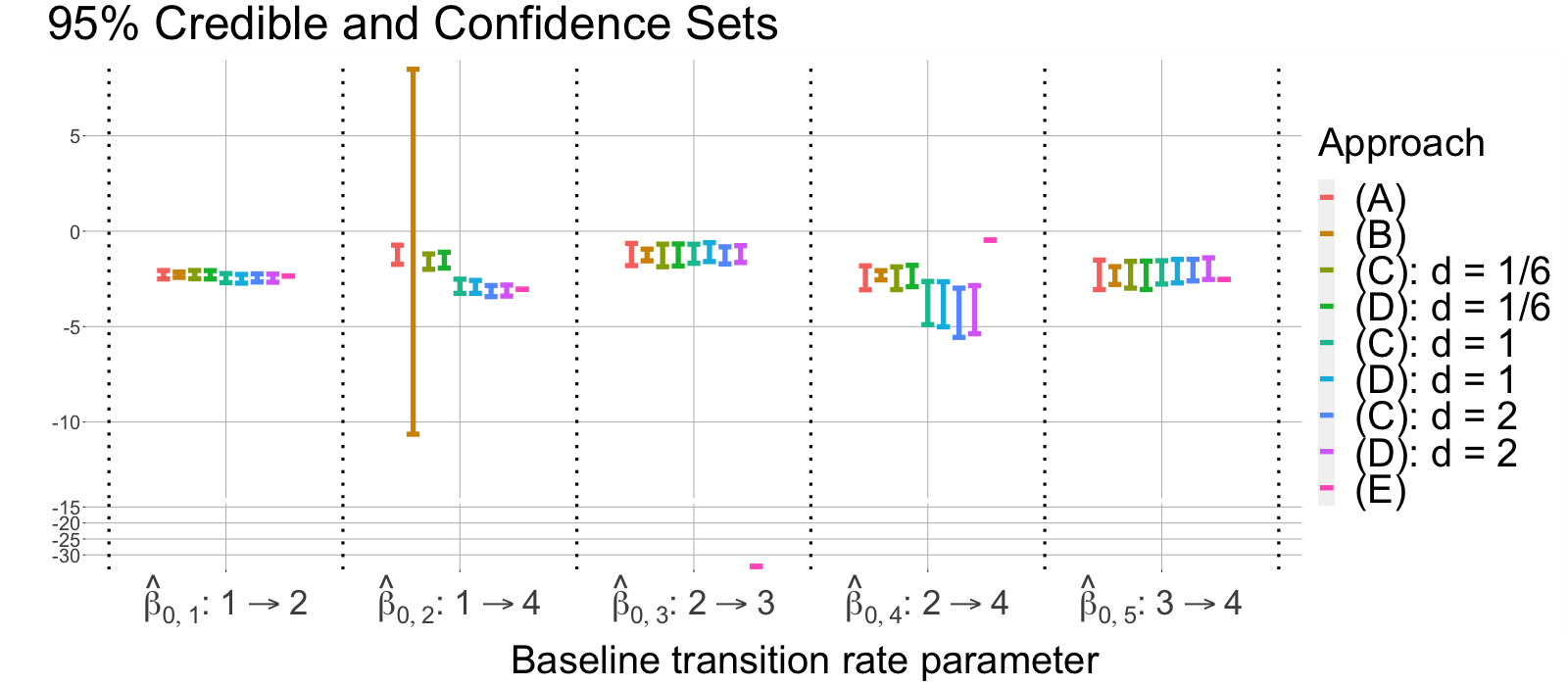}
\caption{\footnotesize For each parameter (i.e., within the dotted vertical lines), the vertical intervals ordered left to right correspond to the ``Approach'' ordered top to bottom. For approaches (A) and (C), highest posterior density credible intervals are reported for the .95 level, for the fitted baseline coefficients, $\beta_{0,j}$, as in equation (\ref{eq:rates_CAV}), for all transitions.  For approaches (B) and (D), .95 level confidence intervals based on the Hessian are reported, as given by the \texttt{nhm} \citep{titmanPackage} and \texttt{msm} packages \citep[][]{jackson2011multi}, respectively. Due to the lack of availability of an approach for constructing a confidence interval for approach (E), only point estimates are provided.}\label{fig:cavRealResults}
\end{figure}

In any case, the .95 level credible or confidence sets for the baseline rate parameters -- fitted by all five approaches and for each $d \in \{1/6,1,2\}$ -- are presented in Figure \ref{fig:cavRealResults}.  Differences are observed in the intervals [across the estimation approaches] most prominently between the numerical ODE solver approaches versus the piecewise-homogeneous approximations for lower resolutions, $d$, and approach (E). Comparing numerical approaches (A) and (B), the confidence intervals of (B) for the baseline parameters tend to be narrower, with the exception of $\beta_{0,2}$. This wide confidence set phenomenon for approach (B) is also seen in the confidence sets for the initial state parameters (provided in the supplemental materials). For completeness, Table \ref{tab:cav_time} outlines the computation speed for the likelihood evaluations using the different approaches. The same trend as in the simulation study of Section \ref{sec:simulation} manifests.  Finally, there is one special case (that we are aware of) in which the piecewise-homogeneous approach (for some $d > 0$) may be a reasonable approximation to the numerical ODE solution, for fitting a time-inhomogeneous HMM with a continuously evolving rate matrix; this special case is discussed in the next section.
\nocite{xu2021use}

\begin{table}[H]
\centering\spacingset{1}\footnotesize
\begin{tabular}{l c c c c }
\hline
approach & (A) & (C): $d=1/6$ & (C): $d=1$  & (C): $d=2$  \\
likelihood evaluation time (seconds) & 1.80 & 1.94 & .59 & .45\\
\hline
\end{tabular}
\caption{\footnotesize Displayed are the times to compute the likelihood function, for each approach evaluated at the posterior means associated with approach (A).}
\label{tab:cav_time}
\end{table}

\subsubsection{Mice sleep study data}\label{mice_data}
The final real data application studied is from the mice sleep study data described in Section \ref{subsec:eeg_mice}. Unlike the irregular observation times in the CAV data, the mice sleep study data are processed such that observation times are equally spaced. This is typically the type of scenario in which a discrete-time (and possibly time-inhomogeneous) Markov model would be appropriate, and that is exactly what the piecewise-homogeneous approach reduces to if the discontinuity points of $g_{d}(\cdot)$ align with the regularly-spaced grid of observed time instances.  If it is suspected, however, that the transition rates evolve continuously in time, then a continuous-time Markov model becomes necessary for proper identification of the transition rates as continuous functions of time (e.g., as log-linear functions of a time-based covariate).  Nonetheless, if the regularly-spaced grid of observed time instances is at a relatively high resolution (i.e., corresponding to relatively small $d > 0$) with respect to the rate of change of the transition rates over time, then -- in the sense of a first-order Taylor approximation -- the piecewise-homogeneous approximation should yield similar inferences as that of the numerical ODE solution.  This follows because the observed time-stamps $t_{i,k}$, for $k \in \{1,\dots,n_{i}\}$ and $i \in \{1,\dots,n\}$, in a data set will {\em not} be distorted (i.e., rounded by some modulus) by the piecewise-homogeneous approach due to the existence of $d > 0$ such that $g_{d}(t_{i,k}) = t_{i,k}$, for all $i,k$, as illustrated by Lemma \ref{lemma:baseline} with $c = 0$. 

We compare -- when $c = 0$ and $d > 0$ is small -- the piecewise-homogeneous and numerical ODE solution approaches to time-inhomogeneous HMM-based inference on a real data set from a mice sleep study. All state transitions are allowed, as specified by the rate matrix presented next.  An additional complication, however, is that the states 1: ``IS'', 2: ``NREM'', and 3: ``REM'' are not an exhaustive list of sleep-states that have been identified by sleep researchers -- they are just a simplified list of sleep-states of interest for this illustration.  That being so, we must define an additional auxiliary state 4 as a catchall state to allow us to properly identify states 1, 2, and 3.  For $t > 0$,
\[
\spacingset{1}
\vb*{Q}(t) = \mqty( -q_1 - q_2-q_3 & q_1 & q_2 & q_3\\
                               q_4 & -q_4 - q_5-q_6 & q_5 & q_6\\
                               q_7 & q_8 & -q_7-q_8-q_9 & q_9\\
                               q_{10} & q_{11} & q_{12} & -q_{10} - q_{11} - q_{12})
\]
where each $q_{j}$, for $j \in \{1,\dots,12\}$, is expressed as 
\begin{equation}\label{eq:rates_mice}
q_{j} = \exp(\beta_{0,j} + \beta_{1,j} \cdot t),
\end{equation}
where $\beta_{0,j}$ and $\beta_{1,j}$ are unknown rate-specific coefficient parameters, for $j \in \{1,\dots,12\}$.

The probability mass function for the observed state labels $Y_{i,k}$ conditional on the true state $s_{i,k}$ is given by the rows of the table:
\begin{center}
\spacingset{1}
\begin{tabular}{c c | c c c c}
& & & $y_{i,k}$ : observed state & \\
& & 1 & 2 & 3 & 4 \\
\hline
& 1 & $1-p_1-p_2$ & $p_1$ & $p_2$ & 0 \\
$s_{i,k}$ : true state & 2 & $p_3$ & $1-p_3-p_4$ & $p_4$ & 0 \\
& 3  & $p_5$ & $p_6$ & $1-p_5 - p_6$ & 0 \\
& 4 & 0 & 0 & 0 & 1 \\
\end{tabular}
\end{center}
Note that the auxiliary state 4 is never actually observed, but is included in the probability mass function for completeness.  Next, we adopt, from \cite{langrock2013combining}, a state-conditional Dirichlet density function for the observed relative powers, denoted $\vb*{X}_{i,k} := (X_{\delta,i,k}, X_{\theta,i,k}, X_{\alpha,i,k}, X_{\beta,i,k})$, of the $\delta$, $\theta$, $\alpha$, and $\beta$ bands, respectively, for all $k \in \{1,\dots,n_{i}\}$ and $i \in \{1,\dots,n\}$:
\[
f_{\lambda}(\vb*{x}_{i,k} \mid s_{i,k} = s) = \frac{\Gamma(\lambda^{(s)}_{\delta} + \lambda^{(s)}_{\theta} + \lambda^{(s)}_{\alpha} + \lambda^{(s)}_{\beta})}{\Gamma(\lambda^{(s)}_{\delta})\Gamma(\lambda^{(s)}_{\theta})\Gamma(\lambda^{(s)}_{\alpha})\Gamma(\lambda^{(s)}_{\beta})} \cdot x_{\delta,i,k}^{\lambda^{(s)}_{\delta} - 1} \cdot x_{\theta,i,k}^{\lambda^{(s)}_{\theta} - 1} \cdot x_{\alpha,i,k}^{\lambda^{(s)}_{\alpha} - 1} \cdot x_{\beta,i,k}^{\lambda^{(s)}_{\beta} - 1},
\]
where $s \in \{1,2,3,4\}$, $\lambda^{(s)}_{\delta}, \lambda^{(s)}_{\theta}, \lambda^{(s)}_{\alpha}, \lambda^{(s)}_{\beta} > 0$, $x_{\delta,i,k}, x_{\theta,i,k}, x_{\alpha,i,k}, x_{\beta,i,k} \in [0,1]$, and $x_{\delta,i,k} + x_{\theta,i,k} + x_{\alpha,i,k} + x_{\beta,i,k} = 1$.

\begin{figure}[H]
\spacingset{1}
\centering
\includegraphics[scale =.275]{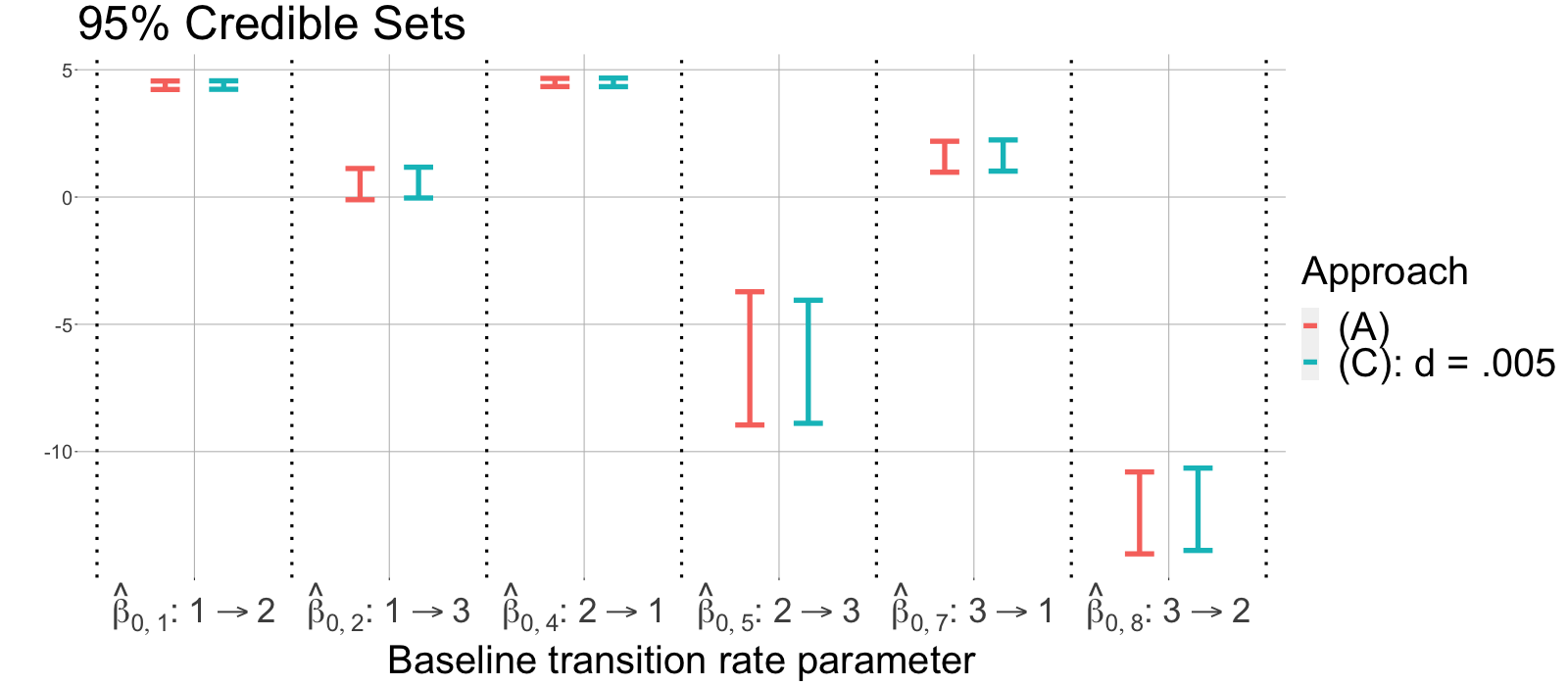}
\includegraphics[scale =.275]{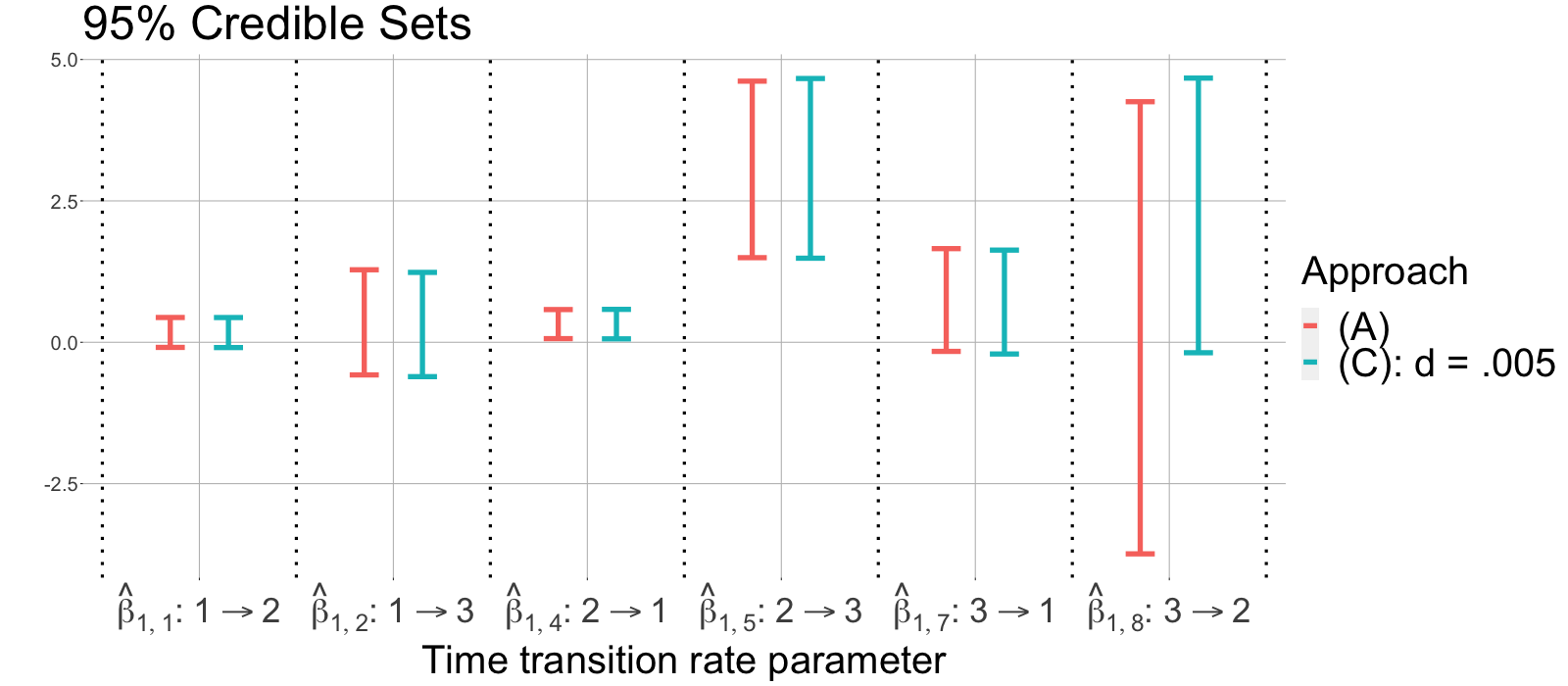}
\caption{\footnotesize For each parameter (i.e., within the dotted vertical lines), the vertical intervals ordered left to right correspond to the ``Approach'' ordered top to bottom. The highest posterior density credible intervals at the .95 level are reported for the fitted coefficients in equation (\ref{eq:rates_mice}), for transitions of interest (i.e., excluding transitions to/from the auxiliary state 4).  The value $d = .005$ is chosen so that $g_{d}(\cdot)$ aligns with the 5-second time increments in the data set (i.e., $g_{d}(t) = t$).}\label{fig:miceResults}
\end{figure}

Figure \ref{fig:miceResults} presents the .95 level credible sets for the rate parameters fitted by approaches (A) and (C) specified in Section \ref{subsec:compete}.  For the piecewise-homogeneous approach, (C), we set $d = .005$ so that $g_{d}(\cdot)$ aligns with the 5-second time increments in the data set (i.e., $g_{d}(t) = t$) scaled by a factor of 1/1000. We find that the credible set intervals always overlap across the two approaches for both the baseline and time transition rate coefficients. This confirms the intuition that for sufficiently small $d>0$, if $g_{d}(t) = t$,  both the numerical ODE solution and matrix-exponential-based solution lead to the same .95 level credible sets. Additionally, we find that, consistent with the belief of sleep researchers, the NREM $\to$ REM transition is non-constant over time. Hence, the piecewise-homogeneous approach is likely a reasonable first-order approximation (based on modeling the rates as log-linear functions of time). Lastly, Table \ref{tab:ecog_time} provides the computation speed for the likelihood evaluations using the two approaches.

\begin{table}[H]
\centering\spacingset{1}\footnotesize
\begin{tabular}{l c c }
\hline
approach & (A) & (C): $d=.005$ \\
likelihood evaluation time (seconds) & 9.40 & 2.70\\
\hline
\end{tabular}
\caption{\footnotesize Displayed are the times to compute the likelihood function, for each approach evaluated at the posterior means associated with approach (A).}
\label{tab:ecog_time}
\end{table}

\section{Concluding remarks}\label{sec:conclusion}
Continuous-time multistate Markov models (HMMs or otherwise) are widely used in data analysis applications involving processes that evolve over time, and generally speaking it is artificial to assume that a naturally occurring process would either exhibit rates of evolution that change in an arbitrarily regular piecewise-constant fashion, or that all observation times correspond to state transition times. Although these assumptions motivate pragmatic approximations for fitting a time-inhomogeneous, continuous-time Markov model, we demonstrate that such approximations may lead to non-negligible biases in estimation. Furthermore, we demonstrate that there exist exact numerical-integration solutions. Moreover, these solutions are computationally more efficient than the approximations, and the computational tools to implement them are readily and freely available. Given the comprehensive list of approaches tested in this analysis, we find that the numerical-integration approaches, based on either Bayesian (approach (A)) or Frequentist (approach (B)) computations, lead to the most consistent estimation of the standard parametric structures that we consider in the simulation study in Section \ref{sec:simulation} and real data analyses in Section \ref{sec:real_data}. Lastly, we note that the numerical integration can be implemented via the Riemann integral (by way of \texttt{deSolve} and/or \texttt{nhm}) or the product integral (by way of \texttt{prodint}).

\section{Appendix}
\subsection{Proofs}\label{sec:proofs}

{\noindent\em Proof of Lemma \ref{lemma:slope}}. 
Let $\vb*{y} := (y_{1}, \dots, y_{n})^{T}$ so that $\vb*{y} = \vb*{W\beta}$, where $\vb*{\beta} := (\beta_{0}, \beta_{1})^{T}$, and
\[
\spacingset{1}
\vb*{W} := \mqty(1 & t_1 \\ \vdots & \vdots \\ 1 & t_n).
\]
Then the least squares projection coefficients in (\ref{eq:projection_coefs}) are obtained from the expression 
\[
\vb*{b}_{n} = (\vb*{X}^{T}\vb*{X})^{-1}\vb*{X}^{T}\vb*{y} = (\vb*{X}^{T}\vb*{X})^{-1}\vb*{X}^{T}\vb*{W\beta},
\]
where $\vb*{b}_{n} := (b_{0,n}, b_{1,n})^{T}$, and
\[
\spacingset{1}
\vb*{X} := \mqty(1 & g_{d}(t_1) \\ \vdots & \vdots \\ 1 & g_{d}(t_n)).
\]

Next, observe that for any $i \in \{1,\dots,n\}$,
\[
t_{i} - \overline{t}_{n} = t_{i} - g_{d}(t_{i}) + g_{d}(t_{i}) - \frac{1}{n}\sum_{k=1}^{n}g_{d}(t_{k}) + \frac{1}{n}\sum_{k=1}^{n}\{g_{d}(t_{k}) - t_{k}\}. 
\]
Multiplying both sides by $\{g_{d}(t_{i}) - \frac{1}{n}\sum_{k=1}^{n}g_{d}(t_{k})\}$, summing over $i \in \{1,\dots,n\}$, and dividing by $A_{n} = \sum_{i=1}^{n}\{g_{d}(t_{i}) - \frac{1}{n}\sum_{k=1}^{n}g_{d}(t_{k})\}^{2} > 0$, for sufficiently large $n$ (by assumption), gives
\[
\frac{ \sum_{i=1}^{n}(t_{i} - \overline{t}_{n})\{g_{d}(t_{i})  - \frac{1}{n}\sum_{k=1}^{n}g_{d}(t_{k}) \}  }{  A_{n} }  =  \frac{ \sum_{i=1}^{n}\{t_{i} - g_{d}(t_{i}) \}\{g_{d}(t_{i})  - \frac{1}{n}\sum_{k=1}^{n}g_{d}(t_{k}) \} }{  A_{n}  }  + 1 + 0.
\]
Then,
\[
\spacingset{1}
\begin{split}
\bigg| \frac{ \sum_{i=1}^{n}(t_{i} - \overline{t}_{n})\{g_{d}(t_{i})  - \frac{1}{n}\sum_{k=1}^{n}g_{d}(t_{k}) \}  }{  A_{n}  } - 1 \bigg| & = \bigg| \frac{ \sum_{i=1}^{n}\{t_{i} - g_{d}(t_{i}) \}\{ g_{d}(t_{i})  - \frac{1}{n}\sum_{k=1}^{n}g_{d}(t_{k}) \} }{  A_{n}  }  \bigg| \\
& \le \frac{ \sum_{i=1}^{n} d \cdot | g_{d}(t_{i})  - \frac{1}{n}\sum_{k=1}^{n}g_{d}(t_{k}) |}{  A_{n} } \\
& = d \cdot \frac{\|v\|_{1}}{\|v\|_{2}^{2}}\\
& \le d \cdot \frac{\sqrt{n}\|v\|_{2}}{\|v\|_{2}^{2}}\\
& \le \frac{d}{(\frac{1}{n}A_{n})^{\frac{1}{2}}},
\end{split}
\]
where $v \in \mathbb{R}^{n}$ is defined by $v_{i} := g_{d}(t_{i}) - \frac{1}{n}\sum_{k=1}^{n}g_{d}(t_{k})$ for $i \in \{1,\dots,n\}$.  Since $d > 0$ is fixed, the proof is completed by applying the assumption that $\lim_{n\to\infty}\frac{1}{n}A_{n} = \infty$.
\hfill $\square$\\

{\noindent\em Proof of Lemma \ref{lemma:baseline}}. 
From the first expression in (\ref{eq:projection_coefs}) and since, by assumption, $t_{i} = g_{d}(t_{i}) + c$,
\begin{align*}
\frac{b_{0,n} - \beta_{0}}{\beta_{1}} & = \frac{\overline{t}_{n}\{\sum_{i=1}^{n}g^{2}_d(t_{i})\} - \frac{1}{n}\sum_{i,j}t_{i}\cdot g_{d}(t_{i})g_{d}(t_{j}) }{ A_{n} } \\
& = \frac{\overline{t}_{n}\{\sum_{i=1}^{n}g^{2}_d(t_{i})\} - \frac{1}{n}\sum_{i,j} \{g_{d}^{2}(t_{i}) + c\cdot g_{d}(t_{i})\} g_{d}(t_{j}) }{ A_{n} } \\
& = \frac{\overline{t}_{n}\{\sum_{i=1}^{n}g^{2}_d(t_{i})\} - \frac{1}{n}\{\sum_{j=1}^{n} g_{d}(t_{j})\} \sum_{i=1}^{n} \{g_{d}^{2}(t_{i}) + c \cdot g_{d}(t_{i})\} }{ A_{n} } \\
& = \frac{\{\overline{t}_{n} - \frac{1}{n}\sum_{j=1}^{n}g_{d}(t_{j})\} \{\sum_{i=1}^{n}g^{2}_d(t_{i})\} - \frac{c}{n}\{\sum_{i=1}^{n}g_{d}(t_{i})\}^{2} }{ A_{n} } \\
& = c\cdot \frac{\{\sum_{i=1}^{n}g^{2}_d(t_{i})\} - \frac{1}{n}\{\sum_{i=1}^{n}g_{d}(t_{i})\}^{2} }{ A_{n} } \\
& = c.
\end{align*}
\hfill $\square$

\subsection{Transition rate parameter estimation via the Nelson-Aalen estimator}\label{sec:ajEst}
Recall that the transition rate matrix, $\vb*{Q}(t)$, from Sections \ref{sec:simulation} and \ref{subsec:cavReal} has the form
\[
\spacingset{1}
\vb*{Q}(t) = \mqty(-q_1 - q_2 & q_1 & 0 & q_2\\
                              0 & -q_3 - q_4 & q_3 & q_4\\
                              0 & 0 & -q_5 & q_5\\
                              0 & 0 & 0 & 0),
\]
where each $q_j$, $j = 1,2,\hdots, 5$, has the following form
$$q_j = q_j(t\mid \text{sex})= \exp(\beta_{0,j} + \beta_{1,j}\cdot t + \beta_{2,j}\cdot \text{sex}),$$ with $t \geq 0$ and $\text{sex} \in \{0,1\}$. The Nelson-Aalen estimator estimates the \textit{cumulative} transition rate function \citep{aalen1978empirical, borgan1997three}. Let $A_j(t \mid \text{sex})$ be the cumulative transition rate function defined as
    $$A_j(t\mid \text{sex}) := \int_0^t q_j(s\mid \text{sex}) \; ds,$$
    and let $\hat{A}_j(t\mid \text{sex})$ denote the Nelson-Aalen estimator. The \texttt{aalen\_johansen} function provides $\hat{A}_j(t\mid \text{sex})$, for each observed transition time, $t \geq 0$, and for $j=1,2,\hdots,5$. Then, in order to retrieve estimates for $\beta_{0,j}$, $\beta_{1,j}$, and $\beta_{2,j}$, $\forall j$, define
    \begin{align*}
        g(\beta_{0,j}, \beta_{1,j}, \beta_{2,j} \mid t, \text{sex}) &:= \int_0^t q_j(s\mid \text{sex}) \; ds\\
        &= \frac{\exp(\beta_{0,j} + \beta_{1,j}\cdot t + \beta_{2,j}\cdot \text{sex}) - \exp(\beta_{0,j} + \beta_{2,j}\cdot \text{sex})}{\beta_{1,j}} ,
    \end{align*}
    and use the $\ell_1$ loss to formulate the parameter estimates as
    $$\widehat{\vb*{\beta}}_j^T = \mqty(\hat{\beta}_{0,j}\\ \hat{\beta}_{1,j}\\ \hat{\beta}_{2,j}) = \argmin{\beta_{0,j}, \beta_{1,j}, \beta_{2,j}} \sum_t \qty|g(\beta_{0,j}, \beta_{1,j}, \beta_{2,j} \mid t, \text{sex}) - \hat{A}_j(t\mid \text{sex})|.$$

\if0\blind
{
\section{Acknowledgements}

Research reported in this publication was supported by the National Heart, Lung, and Blood Institute of the National Institutes of Health under Award Number R56HL155373. The content is solely the responsibility of the authors and does not necessarily represent the official views of the National Institutes of Health. Special thanks is given to Dr. Andrew Titman for his prompt and thorough assistance with implementing the \texttt{nhm} package in our data analyses.
}\fi


{
\spacingset{1}
\bibliographystyle{agsm}
\bibliography{references}
}
\end{document}